\documentclass[journal]{IEEEtran}
\usepackage{amsmath,amsfonts}
\usepackage{algorithmic}
\usepackage{algorithm}
\usepackage{array}
\usepackage[caption=false]{subfig}
\usepackage{textcomp}
\usepackage{stfloats}
\usepackage{url}
\usepackage{verbatim}
\usepackage{graphicx}
\usepackage{cite}
\usepackage{breqn}
\usepackage{amssymb}
\usepackage{placeins}
\begin{document}

\title{Automatic Attenuation Control for Mitigating Photon-Counting Saturation in SPAD-based Optical Wireless Communications}

\author{Chen Wang, Zhiyong Xu, Jingyuan Wang, Jianhua Li, Weifeng Mou, Huatao Zhu
\thanks{This work was supported by National Natural Science Foundation of China (Grant No. 62271502, 62171463, 62471472) Hubei Provincial Natural Science Foundation of China (Grant No. 2026AFA101), and Natural Science Foundation of Jiangsu Province (Grant No. BK20231486). (\textit{Corresponding author: Jingyuan Wang}) 

Chen Wang, Weifeng Mou, Huatao Zhu are with the Information Support Force Engineering University, Wuhan 430010, China (e-mail: 0910210239@njust.edu.cn; weifengmou@126.com; zhuhuatao2008@163.com).

Zhiyong Xu, Jingyuan Wang,  and Jianhua Li are with the College of Communications Engineering, Army Engineering University of PLA, Nanjing 210007, China (e-mail: njxzy123@163.com; 13813975111@163.com; 18021528752@163.com).
}
\thanks{Manuscript received ; revised }}

\markboth{Journal of \LaTeX\ Class Files,~Vol.~xx, No.~xx, May~2026}%
{Shell \MakeLowercase{\textit{et al.}}: A Sample Article Using IEEEtran.cls for IEEE Journals}

\IEEEpubid{}

\maketitle

\begin{abstract}
Single-photon avalanche diodes (SPADs) have emerged as a promising candidate for optical wireless communication (OWC) owing to their ultra-high sensitivity and single-photon detection capability. However, under strong background radiation or high signal power, SPAD-based receivers suffer from photon-counting saturation, which severely degrades communication performance. To address this challenge, this paper introduces an automatic attenuation control (AAC) technique that dynamically optimizes the incident optical intensity to mitigate saturation effects. We develop a comprehensive analytical model for the SPAD-based OWC system, incorporating the influence of dead time and the lack of photon-number resolution. Based on this model, a convex optimization-based AAC algorithm is proposed to maximize the achievable rate in real time. Furthermore, a low-complexity AAC algorithm is devised using a closed-form trigger probability criterion, reducing computational complexity by two orders of magnitude. Numerical results demonstrate that the proposed AAC technique significantly improves both the achievable rate and symbol error rate across a wide range of background conditions, providing an efficient solution to enhance the dynamic range of photon-counting receivers. 
\end{abstract}


\section{Introduction}
\IEEEPARstart{I}{n}  recent years, optical wireless communication (OWC) has garnered significant attention as a promising complement to conventional radio frequency (RF) systems \cite{ref1,ref2}. OWC offers several advantages over RF, including larger bandwidth, lower transmit power, and reduced antenna aperture. Conventional OWC receivers typically employ positive-intrinsic-negative (PIN) photodiodes and avalanche photodiodes (APDs). However, a major limitation of PIN photodiodes is their low gain, which renders the received signal susceptible to domination by thermal noise under extremely low illumination conditions. Although APDs achieve improved sensitivity through internal gain, thereby mitigating thermal noise, the stochastic nature of the avalanche multiplication process introduces excess noise that depends on the gain, thus restricting the maximum usable APD gain \cite{ref3}.

In long-distance photon-starving applications, optical signals may fall below the sensitivity threshold of conventional receivers \cite{ref4}. In such scenarios, single-photon avalanche diodes (SPADs) offer a more suitable alternative. SPADs provide extremely high internal gain, effectively overcoming thermal noise and enabling detection at the single-photon level \cite{ref5}. Owing to their exceptional sensitivity and substantial gain, SPADs have accelerated progress in a variety of applications. These detectors can approach the quantum-limited sensitivity for weak optical signals and have thus attracted considerable interest in OWC researches \cite{ref6, ref7, ref8, ref9}.

In conventional APD receivers, automatic gain control (AGC) is widely used to regulate the avalanche current at an appropriate level \cite{ref10}. In contrast, due to their single-photon sensitivity, SPADs inherently provide avalanche gains exceeding $10^6$, thereby eliminating concerns related to insufficient avalanche gain \cite{ref11}. However, under conditions involving background radiation or non-negligible extinction ratio (ER), SPADs are prone to saturation. In intensity-modulated systems, background photons and extinction photons elevate the photon counts across all signal constellation points. Given the limited count rate, increased incident photon flux can drive the output photon counts into saturation, named photon-counting saturation. This, in turn, reduces the distinguishable photon counts gap between different constellation points. 

Furthermore, the arrival of weak optical photons follows a Poisson process, whose variance is proportional to the mean photon arrival rate. As the incident optical intensity increases, so does the noise intensity. The combined effect of diminished photon counts gap and elevated noise intensity degrades the communication performance, particularly under strong background radiation or high ER conditions. Therefore, introducing an optical attenuator after the receiving aperture to regulate the incident optical power is essential for improving error performance under such saturation-prone conditions. Corresponding to the AGC technique used in APDs, an automatic attenuation control (AAC) mechanism is critically needed in photon-counting receivers to mitigate saturation.

\begin{figure}[!t]
\centering
\includegraphics[width=3.3in]{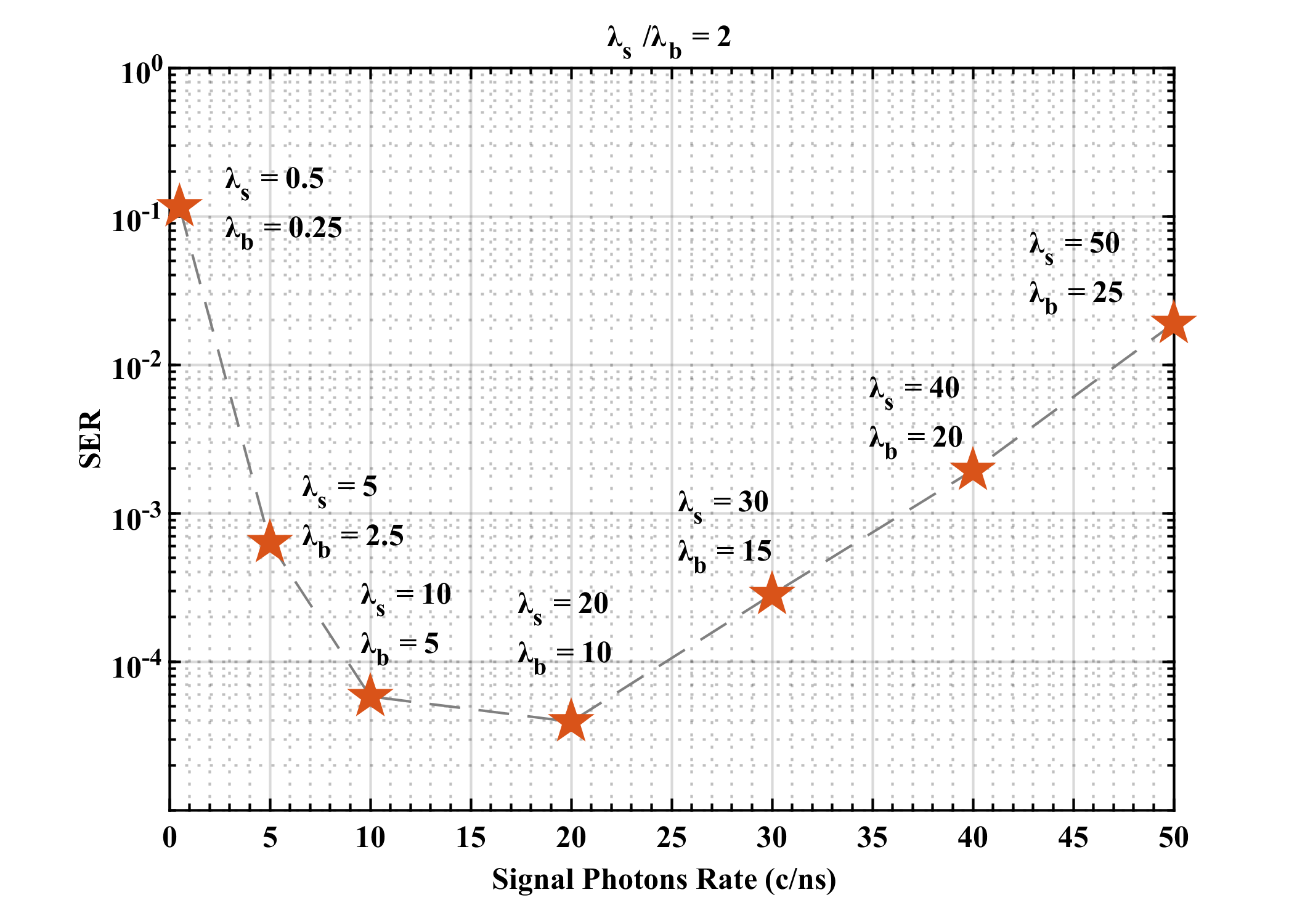}
\caption{SER versus incident optical intensity for a fixed signal-to-background ratio.}
\label{fig_1}
\end{figure}

Benefiting from single-photon sensitivity, photon-counting receivers can operate near the quantum limit \cite{ref12}. Under such conditions, in addition to the Poisson-distributed signal photons, the Poisson noise arising from background radiation becomes significant and cannot be neglected. For clarity, consider an On-Off keying (OOK) OWC system as an example. As the intensities of the signal and background radiation increase, the variances of the photon counts for both "0" and "1" symbols also rise. This produces two counteracting effects on error performance: (1) the increase in signal intensity widens the photon counts gap between symbols, improving symbol error rate (SER); (2) the increased variance elevates the probability of errors, degrading SER. 

Based on a theoretical model established in our prior work \cite{ref13}, Fig. \ref{fig_1} presents simulation results that illustrate these opposing trends. It shows the SER under a fixed signal-to-background ratio (SBR, ${{{\lambda _{\rm{s}}}} \mathord{\left/{\vphantom {{{\lambda _{\rm{s}}}} {{\lambda _{\rm{b}}}}}} \right.\kern-\nulldelimiterspace} {{\lambda _{\rm{b}}}}} = 2$), where SER initially decreases as the incident signal photon rate increases from $0 \!\sim\! 20$ counts/nanosecond ($\rm{c/ns}$), then rises in the range of $20 \!\sim\!50 \hspace{0.2em} \rm{c/ns}$. This behavior confirms that, for a given SBR, an optimal incident optical intensity exists. From a communication standpoint, comprehending the influence of both the SBR and optical intensity on SPAD-based OWC systems is crucial. Accordingly, maintaining the incident optical intensity at its optimal level via AAC is indispensable for practical photon-counting receivers, especially in environments with strong background radiation.

\subsection{Related Works}
Photon-counting receivers typically characterize received signals through discrete photon counts. In such Poisson channels, the conventional additive white Gaussian noise (AWGN) model is not applicable. Early studies often assumed an ideal photon-counting receiver, neglecting dead time effects, and focused on the discrete-time Poisson (DTP) channel model \cite{ref14, ref15}. However, realizing an ideal receiver remains challenging. In practice, dead time introduces nonlinear distortion to the photon counts \cite{ref16}. Specifically, the dead time can inhibit the detection of subsequent photons \cite{ref17}. As a result, photon-counting channels cannot be accurately modeled as DTP channels with merely reduced received signal power \cite{ref18}.

Recent works \cite{ref19, ref20, ref21} have explored time-gated SPAD (TG-SPAD) receivers for long-distance underwater optical wireless communication (UOWC) systems, taking into account blocking loss due to dead time. However, the assumed Poisson statistics do not always provide an accurate approximation. In our previous work \cite{ref22}, we investigated the statistical model of TG-SPAD based on dead time and the lack of photon-number resolution (PNR). Studies \cite{ref23, ref24, ref25} validated TG-SPAD receivers for pulse-position modulation (PPM), incorporating dead time effects and adopting binomial statistics to characterize photon counts, thereby enabling more accurate prediction of system error performance. \cite{ref26, ref27} extended this approach by modeling a TG-SPAD array as a discrete memoryless channel (DMC) for OWC applications.

Beyond non-ideal photon-counting behavior, photon-counting saturation can also degrade error performance. Saturation occurs when the output photon counts approaches its maximum value, which is governed by the symbol duration and dead time. In \cite{ref28}, experimental results demonstrate that under strong background radiation, APD receivers outperform SPAD receivers, as the latter suffer from photon-counting saturation caused by excessive background photons. Studies such as \cite{ref29} and \cite{ref30} suggest that incorporating an attenuation control mechanism in photon-counting receivers can mitigate saturation and enhance dynamic range. However, these works do not elaborate on the implementation details of AAC techniques or related control algorithms. In our prior work \cite{ref31}, we introduced an AAC technique for photon-counting receivers and proposed an automatic control algorithm tailored for OOK modulation. Nevertheless, this approach is limited to OOK and is not directly applicable to higher-order modulation formats such as pulse-amplitude modulation (PAM).

In summary, while AAC techniques for photon-counting receivers have been explored at a qualitative level, quantitative analyses in existing studies remain limited. In practice, controlling the incident optical intensity in an AAC module requires precise values derived from a well-defined automatic control algorithm. Therefore, establishing a comprehensive analytical model for SPAD-based OWC systems employing AAC, applicable across both strong and weak signal regimes, is of critical importance.

\subsection{Our Contributions}
In this paper, we develop a mathematical framework to accurately model the photon-counting saturation in SPAD-based OWC systems. Furthermore, leveraging convex optimization theory, we propose an automatic control algorithm for regulating the incident optical intensity. Although the analysis focuses on SPAD, the proposed approach is applicable to other single-photon detectors with similar photon-counting behaviors, such as superconducting nanowire single-photon detectors (SNSPDs) and photomultiplier tubes (PMTs).

The main contributions are summarized as follows:

{\bf (1) Achievable Rate Model for Non-Ideal Photon-Counting Receivers:} We derive a new closed-form expression for the achievable rate of SPAD-based OWC systems, incorporating the effects of dead time and lack of PNR. This model accurately characterizes the communication performance under both strong and weak background radiation conditions. 

{\bf(2) AAC Algorithm for Mitigating Photon-Counting Saturation:} We propose an AAC algorithm based on convex optimization to counteract photon-counting saturation. This algorithm computes the optimal attenuation coefficient in real time by maximizing the achievable rate. 

{\bf(3) Low-Complexity AAC Algorithm via Trigger Probability Criterion:} To reduce computational complexity, we introduce a closed-form criterion for the attenuation coefficient based on the average trigger probability. This method achieves error performance comparable to the convex optimization. 

\subsection{Organization}
The remainder of this paper is structured as follows. Section~\uppercase\expandafter{\romannumeral2} introduces the AAC technique and provides a formal problem statement. Section~\uppercase\expandafter{\romannumeral3} derives the achievable rate of SPAD-based OWC systems and presents two practical AAC algorithms. Section~\uppercase\expandafter{\romannumeral4} provides numerical results and discussions. Finally, Section~\uppercase\expandafter{\romannumeral5} concludes the paper. Notations used are collated in Table \ref{tab_1}. 

\renewcommand{\arraystretch}{1.2}
\begin{table}[!t]
\centering
\caption{Notations and preliminaries}
\label{tab_1}
\begin{tabular}{|>{\centering\arraybackslash}m{0.20\linewidth}|>{\centering\arraybackslash}m{0.70\linewidth}|} 
\hline
\textbf{Notation} & \textbf{Definition} \\
\end{tabular}
\begin{tabular}{|>{\centering\arraybackslash}m{0.20\linewidth}|>{\raggedright\arraybackslash}m{0.70\linewidth}|} 
\hline
        $p_{\rm{d}}, N_{\rm{A}}$ & Photon detection efficiency, array scale (number of SPAD pixels) \\ \hline
        $T_{\rm{s}}, \tau_{\rm{d}}, \tau_{\rm{g}}, k_{\max}$ & Symbol duration, dead time, gate duration, maximum photon counts \\ \hline
        $\lambda_{\rm{s}}, \lambda_{\rm{b}}, \lambda_{\rm{d}}$ & Signal photon rate, background photon rate, dark carrier rate \\ \hline
        $N(t)$ & Stochastic variable denoting the number of events in the time interval $(0,t)$ \\ \hline
        $M, m$ & PAM modulation order, symbol information \\ \hline
        $p_{\rm{tri}}(\alpha,m)$ & Trigger probability for symbol ``$m$'' \\ \hline
        $p_{m,k}$ & Probability of detecting $k$ photons for symbol ``$m$'' \\ \hline
        $I(Y;X)$ & Mutual information between the transmitted symbol $X$ and the detected count $Y$ \\ \hline
        $P_{\rm{e}},k_{\rm{th}}$ & Symbol error rate, threshold for signal demodulation \\ \hline
\end{tabular}
\end{table}

\section{System Model, Assumptions, and Problem Statement}
\begin{figure}[!t]
\centering
\includegraphics[width=3.3in]{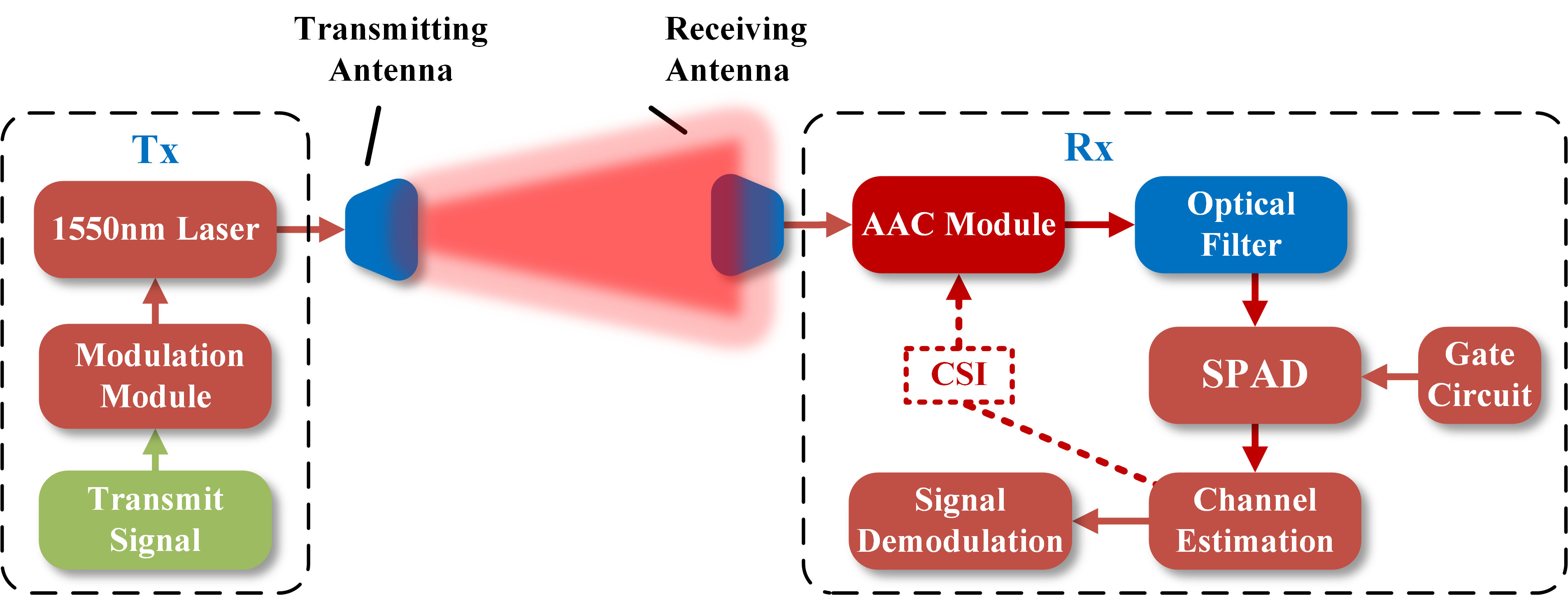}
\caption{Diagram of the proposed scheme using AAC technique in photon-counting receiver.}
\label{fig_2}
\end{figure}

As illustrated in Fig. \ref{fig_2}, we propose an AAC technique for practical photon-counting receivers to mitigate photon-counting saturation. The AAC module operates by dynamically adjusting attenuation coefficient to reduce excessive incident photon flux in real time. The adjustment strategy relies on channel state information (CSI), obtained by detecting pilot symbols. The estimates of the incident signal and background photon rates are fed back to the AAC module. When the incident optical intensity exceeds the optimal level, the AAC module activates to attenuate the incoming light, adjusting the attenuation coefficient to maintain optimal incident intensity. The expression for the optimal attenuation coefficient is:
\begin{equation}\label{eq1}
{\alpha _{{\rm{opt}}}} = \frac{{{\lambda _{{\rm{s\!-\!opt}}}} + {\lambda _{{\rm{b\!-\!opt}}}}}}{{{\lambda _{\rm{s}}} + {\lambda _{\rm{b}}}}}
\end{equation}
where ${\lambda _{{\rm{s\!-\!opt}}}}$  and ${\lambda _{{\rm{b\!-\!opt}}}}$ denote the optimal signal and background photon rates, respectively, that maximize system performance for a fixed SBR.

Photon arrival processes in photon-starved optical links are commonly modeled as Poisson processes, consistent with the statistical nature of weak optical signals \cite{ref32}. Let $N\left( t \right)$ represent the random variable corresponding to the number of photon arrivals within a time interval $\left( {0,t} \right)$. For an ideal single-photon detector (neglecting dead time) with photon detection efficiency (PDE) ${p_{\rm{d}}}$, each incident photon is detected independently with probability ${p_{\rm{d}}}$. Dark counts, occurring at an average rate ${\lambda _{\rm{d}}}$, also contribute to the overall photon counts. The resulting detected photon process remains Poisson-distributed: $N\left( t \right) \!\sim\! \pi \left( {\lambda t} \right)$. Thus, the probability mass function (PMF) of detecting $k$ photons in the time interval $\left( {0,t} \right)$ is given by \cite{ref33}:
\begin{equation}\label{eq2}
P\left\{ {N(t) = k} \right\} = \frac{{{{\left( {\lambda t} \right)}^k}}}{{k!}}{{\rm{e}}^{ - \lambda t}}
\end{equation}
where $\lambda  \ge 0$ denotes the average detected photon rate.

In addition to the desired signal photons, background radiation and dark counts contribute indistinguishable spurious photon counts. To mitigate the inherent randomness of single-photon detection, symbol information is typically recovered by integrating photon counts over the symbol duration. As a result, the continuous optical pulse stream is converted into a sequence of discrete detection events.

Dead time introduces nonlinear distortion in the photon counts, particularly noticeable at high incident optical power. To maintain acceptable communication performance, the maximum photon counts per symbol is typically several dozens. The maximum photon counts ${k_{\max }}$ depends on the dead time ${\tau _{\rm{d}}}$, gate duration ${\tau _{\rm{g}}}$, symbol duration ${T_{\rm{s}}}$ and array scale  ${N_{\rm{A}}}$. For this detection scheme, ${k_{\max}}$ is given by:
\begin{equation}\label{eq3}
{k_{\max }} = {N_{\rm{A}}}\left\lceil {\frac{{{T_{\rm{s}}}}}{{{\tau _{\rm{d}}} + {\tau _{\rm{g}}}}}} \right\rceil
\end{equation}
where $\left\lceil  \cdot  \right\rceil$ denotes the ceiling function.

We then define the trigger probability ${p_{\rm{tri}}}$ as the likelihood of an avalanche event being triggered by an incident photon within a gate duration. Based on (2), this probability is:
\begin{equation}\label{eq4}
{p_{\rm{tri}}} = P\left\{ {N\left( {{\tau _{\rm{g}}}} \right) \ge 1} \right\} = 1 - {{\mathop{\rm e}\nolimits} ^{ - \alpha \lambda {\tau _{\rm{g}}}}}
\end{equation}
where $\alpha $ is the attenuation coefficient.

Assuming the signal photon rate, background photon rate, and dark carrier rate remain constant over a symbol duration, let $X = \left\{ {{\lambda _1}, \ldots ,{\lambda _m}, \ldots ,{\lambda _M}} \right\}$ represent the transmitted symbol constellation for $M$-PAM. Over a symbol duration, the received photon counts provides a probabilistic measure of the incident photon number. The conditional PMF of $Y = k$  given $X = {\lambda _m}$ follows a Binomial distribution \cite{ref13}:
\begin{dmath}\label{eq5}
P\left\{ {Y = k\left| {X = {\lambda _m}} \right.} \right\} 
= \binom{k_{\max}}{k} {p_{\rm{tri}}}^k{\left( {1 - {p_{\rm{tri}}}} \right)^{{k_{\max }} - k}} \\
  = \binom{k_{\max}}{k} {\left( {1 - {{\mathop{\rm e}\nolimits} ^{ - \lambda _m^{{\rm{tot}}}}}} \right)^k}
 \left( {{{\mathop{\rm e}\nolimits} ^{ - \lambda _m^{{\rm{tot}}}}}} \right)^{{k_{\max }} - k}\\
 \triangleq {p_{m,k}}
\end{dmath}
where $\lambda _m^{{\rm{tot}}}$ denotes the total photon arrival rate without AAC, given by $\lambda _m^{{\rm{tot}}} \triangleq \left[ {{p_{\rm{d}}}\left( {{\lambda _m} + {\lambda _{\rm{b}}}} \right) + {\lambda _{\rm{d}}}} \right]{\tau _{\rm{g}}}$. $k \!=\! 0,1, \cdots ,{k_{\max }}$ is a non-negative integer, $\binom{k_{\max}}{k}$ is combinatorial number.

As shown in (\ref{eq5}), a key distinction between the photon-limited OWC and conventional wireless communication is that the received photon counts $Y$ is subject to signal-dependent noise (SDN) arising from the photon-counting process in SPADs, commonly referred to as Poisson shot noise. This shot noise depends on both the signal and background photon rates, whereas the widely used AWGN model accounts only for signal-independent white noise.

{\bf Problem Statement:} Our objective is to maximize the achievable rate of the SPAD-based photon-limited OWC system by optimizing the attenuation coefficient, under the assumption that the signal and background photon rates have been accurately estimated.

\section{Proposed AAC Technique and Algorithms}
We aim to maximize the achievable rate by optimizing the attenuation coefficient, subject to the physical constraints $0 < \alpha \le 1$. This begins with the derivation of an exact expression for the achievable rate under photon-counting receiver, accounting for dead time and the lack of PNR.

\subsection{Achievable Rate and Convexity Analysis}
The achievable rate of the SPAD-based OWC system is derived by evaluating the mutual information between the transmitted symbol $X$ and the received photon counts $Y$, expressed at the top of next page (\ref{eq6}).
\begin{figure*}[!t]
\begin{subequations}\label{eq6}
\begin{align}
I\left( {Y;X} \right) &= H\left( Y \right) - H\left( {Y\left| X \right.} \right) \nonumber\\
&= - \sum\limits_{k = 0}^{{k_{\max }}} {P\left\{ {Y = k} \right\}{{\log }_2}\left[ {P\left\{ {Y=k} \right\}} \right]}  + \sum\limits_{k = 0}^{{k_{\max }}} {\sum\limits_{m = 1}^M {P\left\{ {Y = k,X ={\lambda _m}} \right\}{{\log }_2}\left[ {P\left\{ {Y = k,X = {\lambda _m}} \right\}} \right]} } \label{eq6-1}\\
&= \sum\limits_{k = 0}^{{k_{\max }}} {\sum\limits_{m = 1}^M {\frac{{P\left\{ {Y = k\left| {X = {\lambda _m}} \right.} \right\}}}{M}{{\log }_2}\left[ {\frac{{P\left\{ {Y = k\left| {X = {\lambda _m}} \right.} \right\}}}{M}} \right]} }  \nonumber\\
&\quad - \sum\limits_{k = 0}^{{k_{\max }}} {\left[ {\sum\limits_{m = 1}^M {\frac{{P\left\{ {Y = k\left| {X = {\lambda _m}} \right.} \right\}}}{M}} } \right]{{\log }_2}\left[ {\sum\limits_{m = 1}^M {\frac{{P\left\{ {Y = k\left| {X = {\lambda _m}} \right.} \right\}}}{M}} } \right]}  \label{eq6-2}\\
&= \underbrace {\sum\limits_{k = 0}^{{k_{\max }}} {\overbrace {\sum\limits_{m = 1}^M {\frac{{P\left\{ {Y = k\left| {X = {\lambda _m}} \right.} \right\}}}{M}} }^{ = P\left\{ {Y = k} \right\}}} }_{ = 1}{\log _2}M + \frac{1}{{M\ln 2}}\sum\limits_{k = 0}^{{k_{\max }}} {\sum\limits_{m = 1}^M {P\left\{ {Y = k\left| {X = {\lambda _m}} \right.} \right\}\ln \left[ {P\left\{ {Y = k\left| {X = {\lambda _m}} \right.} \right\}} \right]} } \nonumber\\ 
&\quad- \frac{1}{{M\ln 2}}\sum\limits_{k = 0}^{{k_{\max }}} {\left[ {\sum\limits_{m = 1}^M {P\left\{ {Y = k\left| {X = {\lambda _m}} \right.} \right\}} } \right]} \ln \left[ {\sum\limits_{m = 1}^M {P\left\{ {Y = k\left| {X = {\lambda _m}} \right.} \right\}} } \right]  \label{eq6-3}\\
&= {\log _2}M + \frac{1}{{M\ln 2}}\left\langle {\sum\limits_{k = 0}^{{k_{\max }}} {\sum\limits_{m = 1}^M {P\left\{ {Y = k\left| {X = {\lambda _m}} \right.} \right\}\ln \left[ {P\left\{ {Y = k\left| {X = {\lambda _m}} \right.} \right\}} \right]} } } \right. \nonumber\\
&\hspace{10em} - \left. {\sum\limits_{k = 0}^{{k_{\max }}} {\left[ {\sum\limits_{m = 1}^M {P\left\{ {Y = k\left| {X = {\lambda _m}} \right.} \right\}} } \right]\ln \left[ {\sum\limits_{m = 1}^M {P\left\{ {Y = k\left| {X = {\lambda _m}} \right.} \right\}} } \right]} } \right\rangle  \label{eq6-4}\\
&= {\log _2}M + \frac{1}{{M\ln 2}}\left\langle {\sum\limits_{k = 0}^{{k_{\max }}} {\sum\limits_{m = 1}^M {{p_{m,k}}\ln \left[ {{p_{m,k}}} \right]} }  - \sum\limits_{k = 0}^{{k_{\max }}} {\left[ {\sum\limits_{m = 1}^M {{p_{m,k}}} } \right]\ln \left[ {\sum\limits_{m = 1}^M {{p_{m,k}}} } \right]} } \right\rangle  \label{eq6-5}
\end{align}
\end{subequations}
\vspace{1em}
\hrulefill 
\end{figure*}
${p_{m,k}}$ is a function of $\alpha$, ${\lambda _m}$ and ${\lambda _{\rm{b}}}$; see (\ref{eq5}). Step (\ref{eq6-1}) is based on the definition of mutual information. (\ref{eq6-2}) is obtained by substitution $P\left\{ {X = {\lambda _m}} \right\} = \frac{1}{M}$ and application of the total probability rule; (\ref{eq6-3}) substitutes the Binomial PMF from (\ref{eq5}); and (\ref{eq6-4}) simplifies using $\sum\nolimits_{k = 0}^{{k_{\max }}} {P\left\{ {Y = k} \right\}}  = 1$.

The resulting expression for achievable rate in (\ref{eq6-5}) is mathematically intractable and unsuitable for direct optimization. To proceed, we analyze the convexity of (\ref{eq6-5}) by examining its second-order derivative with respect to $\alpha$ as shown in (\ref{eq7}).
\begin{figure*}[!t]
\begin{align}\label{eq7}
\frac{{d^2}I(Y;X)}{{d{\alpha ^2}}}\!=\!\frac{1}{{M\ln 2}}\sum\limits_{k = 0}^{{k_{\max }}} \sum\limits_{m = 1}^M \binom{k_{\max}}{k}\left\langle {f''_m}\left( \alpha  \right)\left( {1\!+\!\ln \left[ {\frac{{{f_m}\left( \alpha  \right)}}{{g\left( \alpha  \right)}}} \right]} \right)\!+\!{f'_m}\left( \alpha  \right)\left[ {\frac{{{f'_m}\left( \alpha  \right)}}{{{f_m}\left( \alpha  \right)}}\!-\!\frac{{2g'\left( \alpha  \right)}}{{g\left( \alpha  \right)}}} \right]\!-\!{f_m}\left( \alpha  \right)\left[ {\frac{{g''\left( \alpha  \right)}}{{g\left( \alpha  \right)}}\!-\!\frac{{g'{{\left( \alpha  \right)}^2}}}{{g{{\left( \alpha  \right)}^2}}}} \right] \right\rangle 
\end{align}
\vspace{1em}
\hrulefill 
\end{figure*}
${f_m}\left( \alpha \right)$ and $g\left( \alpha \right)$ are components of the Binomial PMF. The first-order and second-order derivatives ${f'_m}\left( \alpha  \right),g'\left( \alpha  \right),{f''_m}\left( \alpha  \right)$ and $g''\left( \alpha  \right)$ are given in closed-form by (\ref{eq8})--(\ref{eq13}):
\begin{align}
{f_m}\left( \alpha  \right) &= {\left( {1 - {{\rm{e}}^{ - \alpha \lambda _m^{{\rm{tot}}}}}} \right)^k}{\left( {{{\rm{e}}^{ - \alpha \lambda _m^{{\rm{tot}}}}}} \right)^{{k_{\max }} - k}} \label{eq8} \\
{f'_m}\left( \alpha  \right)\!&=\!\frac{{\lambda _m^{{\rm{tot}}}{{\left( {{{\rm{e}}^{\alpha \lambda _m^{{\rm{tot}}}}}\!-\!1} \right)}^{k\!-\!1}}}}{{{{\left( {{{\rm{e}}^{\alpha \lambda _m^{{\rm{tot}}}}}} \right)}^{{k_{\max }}}}}}\!\left[ {{k_{\max }}\!+\!{{\rm{e}}^{\alpha \lambda _m^{{\rm{tot}}}}}\left( {k\!-\!{k_{\max }}} \right)} \right] \label{eq9}\\
{f''_m}\left( \alpha  \right) &= \frac{{{{\left( {\lambda _m^{{\rm{tot}}}} \right)}^2}{{\left( {{{\rm{e}}^{\alpha \lambda _m^{{\rm{tot}}}}} - 1} \right)}^{k - 2}}}}{{{{\left( {{{\rm{e}}^{\alpha \lambda _m^{{\rm{tot}}}}}} \right)}^{{k_{\max }}}}}} \label{eq10}\\
& \hspace{2em} \times \left\langle -{k{{\rm{e}}^{\alpha \lambda _m^{{\rm{tot}}}}}} + {{{\left[ {{k_{\max }} + {{\rm{e}}^{\alpha \lambda _m^{{\rm{tot}}}}}\left( {k - {k_{\max }}} \right)} \right]}^2}} \right\rangle \nonumber \\
g\left( \alpha  \right) &= \sum\limits_{m = 1}^M {{f_m}\left( \alpha  \right)} \label{eq11} \\
g'\left( \alpha  \right) &= \sum\limits_{m = 1}^M {{f'_m}\left( \alpha  \right)} \label{eq12} \\
g''\left( \alpha  \right) &= \sum\limits_{m = 1}^M {{f''_m}\left( \alpha  \right)} \label{eq13} 
\end{align}

\begin{figure}[!thb]
\centering
\subfloat[Weak background radiation]{\includegraphics[width=3.3in]{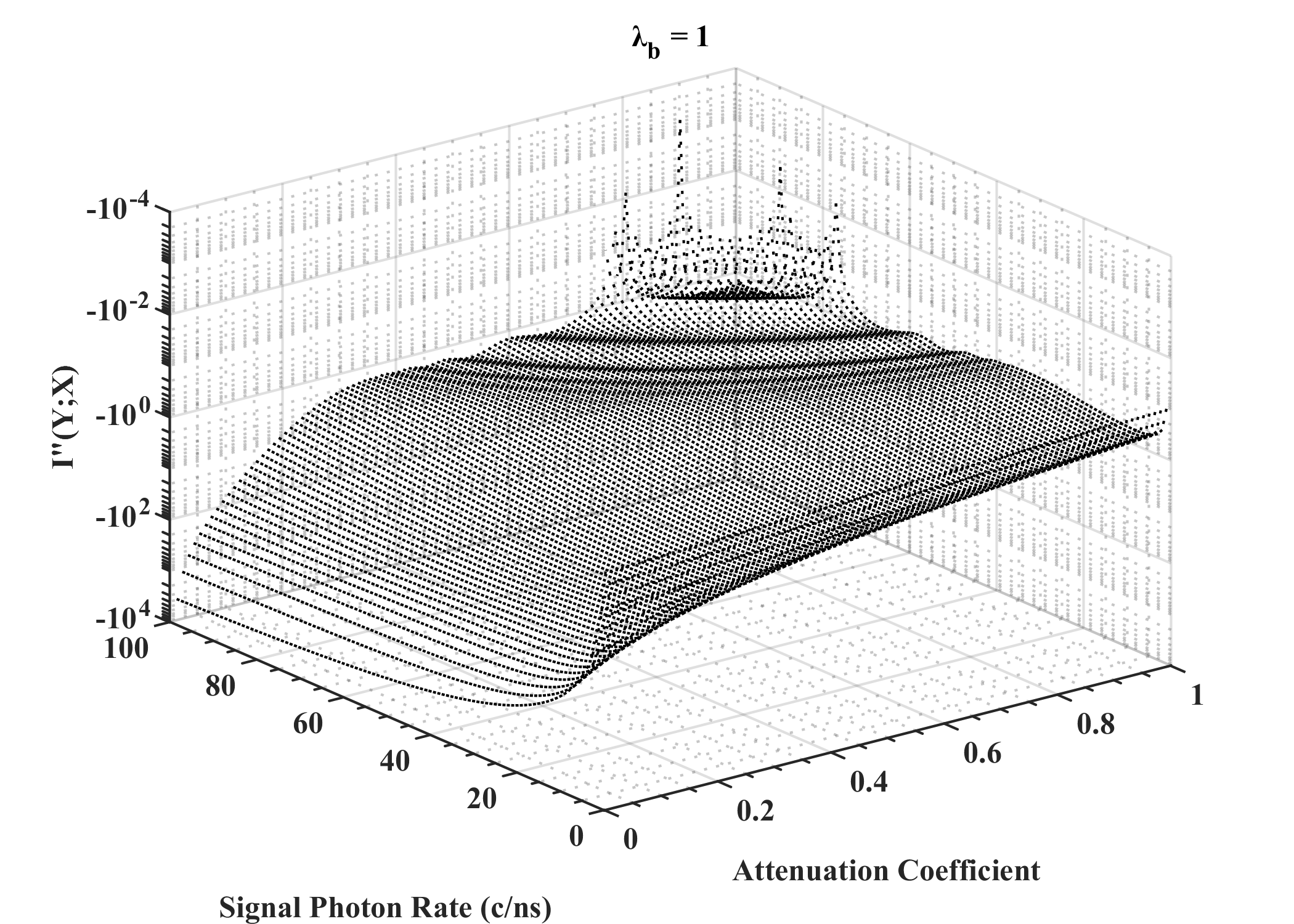}\label{fig_3_1}}
\hfil
\subfloat[Medium background radiation]{\includegraphics[width=3.3in]{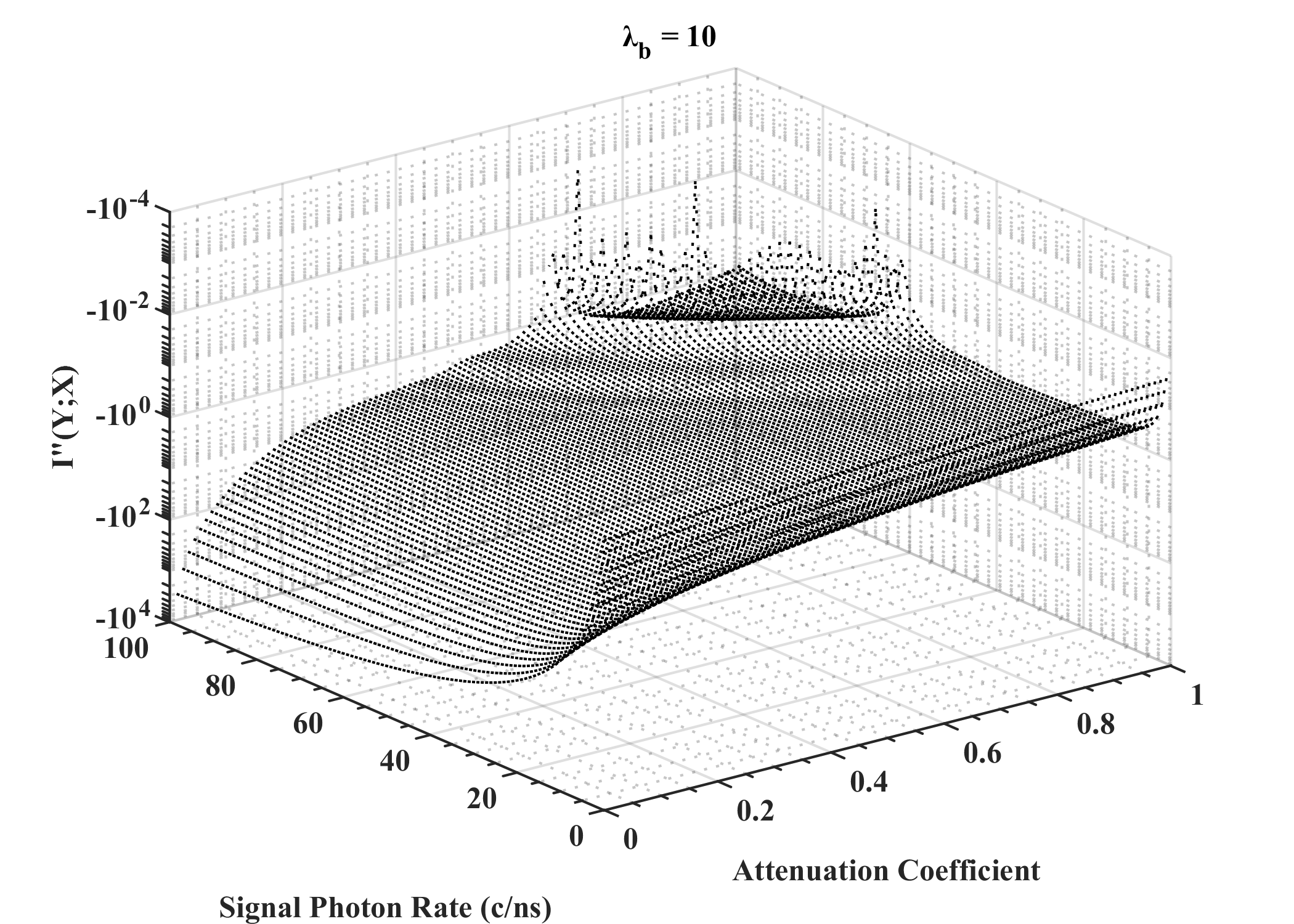}\label{fig_3_2} } 
\hfil
\subfloat[Strong background radiation]{\includegraphics[width=3.3in]{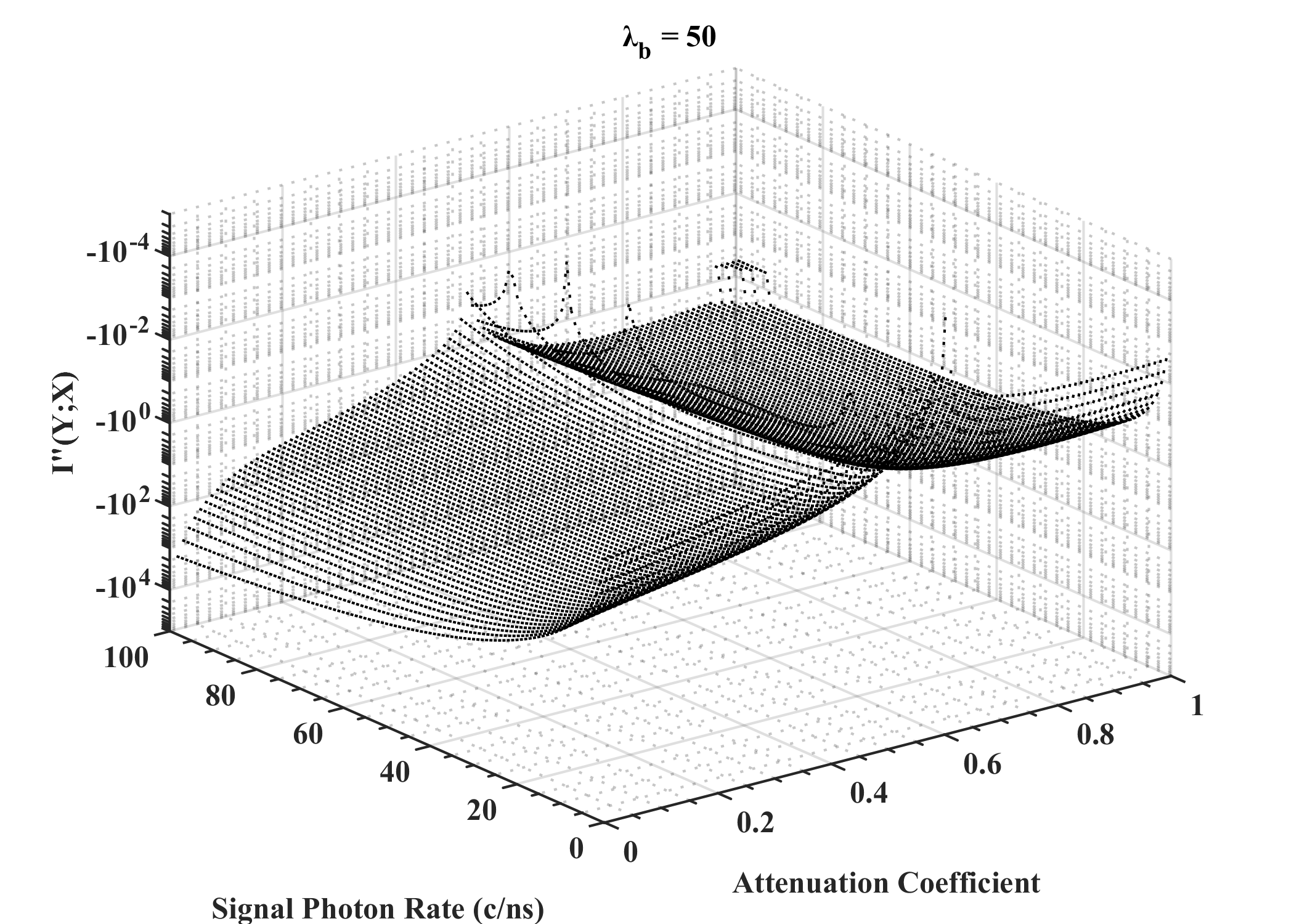}\label{fig_3_3}} 
\hfil
\caption{The contour map of $\frac{d^2 I(Y;X)}{d{\alpha ^2}}$ as a function of attenuation coefficient and signal photon rate.}
\label{fig_3}
\end{figure}

Although (\ref{eq7})--(\ref{eq13}) are provided in closed-form, their analytical tractability remains limited. Therefore, we numerically evaluate the convexity of the achievable rate with respect to the attenuation coefficient. By numerically computing (\ref{eq7}), Fig. \ref{fig_3} illustrates the second-order derivative $\frac{d^2 I(Y;X)}{d{\alpha ^2}}$ under three background radiation levels (weak, medium, and strong), considering peak signal photon rates ranging from $1 \!\sim\! 100 \hspace{0.2em} \rm{c/ns}$. $\frac{d^2 I(Y;X)}{d{\alpha ^2}}$ is initially negative, then increases and converges to zero as $\alpha$ increases. Crucially, it remains non-positive across all considered scenarios, confirming that the achievable rate is concave with respect to $\alpha$.

\subsection{Problem Formulation}
We formulate the optimization problem to maximize the achievable rate as follows:
\begin{subequations}\label{eq14}
\begin{align}
{\bf{P1:}}\mathop {\max }\limits_\alpha  I\left( {Y;X} \right) \label{eq14-1}\\
{\rm{s.t.}} \hspace{0.5em} {\rm{ 0 < }}\alpha  \le 1 \label{eq14-2}
\end{align}
\end{subequations}
Constraint (\ref{eq14-2}) ensures the AAC module only attenuates or leaves the incident light unchanged. Given the concavity of the objective, Problem {\bf P1} can be efficiently solved using convex optimization methods.

\subsection{AAC Algorithms}
As established in Section. III-A, the achievable rate is concave with respect to $\alpha$. Accordingly, we develop a customized optimization algorithm based on the gradient descent method to locate the optimal attenuation coefficient. The detailed procedure is summarized as in Algorithm 1 (Alg. \ref{alg_1}). Using Alg. \ref{alg_1}, the AAC module can dynamically adapt to variations in the background radiation environment. Although Alg. \ref{alg_1} enables adaptive AAC operation, its computational complexity is high due to repeated derivative calculations. 

With a convergence accuracy of $\varepsilon$, the gradient descent method in Alg. \ref{alg_1} requires  $\mathcal{O}\left( {\log \left( {{1 \mathord{\left/ {\vphantom {1 \varepsilon }} \right. \kern-\nulldelimiterspace} \varepsilon }} \right)} \right)$ iterations. In each iteration, the computation involves summation over $M$ signal constellation points and ${k_{\max }}$ possible photon counts, leading to a per-iteration complexity of $\mathcal{O}\left( {M{k_{\max }}}\right)$. Including the cost of invoking the CVX toolbox, the overall complexity of Alg. \ref{alg_1} amounts to $\mathcal{O}\left( {M{k_{\max }}\log \left( {{1 \mathord{\left/ {\vphantom {1 \varepsilon }} \right. \kern-\nulldelimiterspace} \varepsilon }} \right)} \right)$.

To reduce complexity, we propose a simplified algorithm based on a trigger probability criterion. The criterion is as follows:
\begin{align}\label{eq15}
&\frac{1}{M}\sum\limits_{m = 1}^M {{p_{\rm{tri}}}\left( {\alpha ,m} \right)}  \to 0.7 \nonumber\\
 \Rightarrow &\frac{1}{M}\sum\limits_{m = 1}^M {\left( {1 - {{\rm{e}}^{ - \alpha \lambda _m^{{\rm{tot}}}}}} \right) - 0.7}  \to 0\\
 \Rightarrow &\mathop {\min }\limits_\alpha  \left| {\frac{1}{M}\sum\limits_{m = 1}^M {\left( {1 - {{\rm{e}}^{ - \alpha \lambda _m^{{\rm{tot}}}}}} \right) - 0.7} } \right| \nonumber
\end{align}
Since the photon counts variance under Binomial statistics peaks at a trigger probability of $0.5$. We aim to maintain the average trigger probability near $0.7$, which is empirically validated via Alg. \ref{alg_1}. Based on the criterion (\ref{eq15}), we further propose a low-complexity alternative AAC algorithm. The specific steps are outlined in Alg. \ref{alg_2}.

\begin{algorithm} [!thb]
\caption{Gradient-based AAC Algorithm}
\begin{algorithmic}\label{alg_1}
\STATE 
\STATE {\textbf{Output:} Estimated signal and background photon rates $
{\lambda _{{\rm{s\!-\!est}}}},{\lambda _{{\rm{b\!-\!est}}}}$.}
\STATE {\textbf{Output:} Optimal attenuation coefficient ${\alpha _{{\rm{opt}}}}$.}
\STATE \textbf{1:}Initialize ${\alpha ^{\left( 0 \right)}}{\rm{ = 1}}$, iteration index $t = 0$;
\STATE \textbf{2:while} $t\!<\!{T_{\max}}$ and ${\left[ {I{{\left( {Y;X} \right)}^{\left( t \right)}}\!-\!I{{\left( {Y;X} \right)}^{\left( {t - 1} \right)}}} \right]^2} \!>\! \varepsilon $ do
\STATE \textbf{3:}\hspace{0.2cm} Inserting ${\lambda _{{\rm{s\!-\!est}}}}$, ${\lambda _{{\rm{b\!-\!est}}}}$ and ${\alpha ^{\left( t \right)}}$ into (\ref{eq6});
\STATE \textbf{4:}\hspace{0.2cm} 
$t \leftarrow t + 1$;
\STATE \textbf{5:}\hspace{0.2cm} Obtain ${\alpha ^{\left( t \right)}}$, by using the CVX toolbox;
\STATE \textbf{6:end while}
\STATE \textbf{7:return} ${\alpha _{{\rm{opt}}}} = {\alpha ^{\left( t \right)}}$.
\end{algorithmic}
\end{algorithm}

\begin{algorithm} [!thb]
\caption{AAC Algorithm via Trigger Probability Criterion}
\begin{algorithmic}\label{alg_2}
\STATE 
\STATE {\textbf{Output:}Estimated signal and background photon rates $
{\lambda _{{\rm{s\!-\!est}}}},{\lambda _{{\rm{b\!-\!est}}}}$.}
\STATE {\textbf{Output:} Optimal attenuation coefficient ${\alpha _{{\rm{opt}}}}$.}
\STATE \textbf{1:}Initialize ${\alpha ^{\left( 0 \right)}}{\rm{ = 1}}$, iteration index $t = 0$;
\STATE \textbf{2:while} $t\!<\!{T_{\max}}$ and ${\left[ \frac{1}{M}{\sum\nolimits_{m\!=\!1}^M\!{{p_{\rm{tri}}}\left( {\alpha^{(t)},m} \right)}\!-\!0.7} \right]^2}\!>\!\varepsilon $ do
\STATE \textbf{3:}\hspace{0.2cm} Inserting ${\lambda _{{\rm{s\!-\!est}}}}$, ${\lambda _{{\rm{b\!-\!est}}}}$ and ${\alpha ^{\left( t \right)}}$ into (\ref{eq15});
\STATE \textbf{4:}\hspace{0.2cm} 
$t \leftarrow t + 1$;
\STATE \textbf{5:}\hspace{0.2cm} Obtain ${\alpha ^{\left( t \right)}}$, by using Newton–Raphson method;
\STATE \textbf{6:end while}
\STATE \textbf{7:return} ${\alpha _{{\rm{opt}}}} = {\alpha ^{\left( t \right)}}$.
\end{algorithmic}
\end{algorithm}

Since the average trigger probability is twice differentiable and the initial point lies close to the optimal solution. Under the same convergence accuracy $\varepsilon $, the Newton–Raphson method used in Alg. \ref{alg_2} converges within $\mathcal{O}\left( {\log \log \left( {{1 \mathord{\left/ {\vphantom {1 \varepsilon }} \right. \kern-\nulldelimiterspace} \varepsilon }} \right)} \right)$ iterations. Each iteration requires only operations for evaluating relevant PMFs. Thus, the total complexity of Alg. \ref{alg_2} is $\mathcal{O}\left( {M\log \log \left( {{1 \mathord{\left/ {\vphantom {1 \varepsilon }} \right.\kern-\nulldelimiterspace} \varepsilon }} \right)} \right)$. Given that ${k_{\max }}$ typically ranges in several dozens, Alg. \ref{alg_2} reduces the computational complexity by more than two orders of magnitude compared to Alg. \ref{alg_1}.

\subsection{Performance Metrics}
In addition to achievable rate, we evaluate the SER to validate the AAC technique. The SER is derived from the photon counts PMFs as \cite{ref13}:
\begin{dmath}\label{eq16}
{P_{\rm{e}}} = \frac{1}{M}\sum\limits_{m = 1}^{M - 1} {\left\{ {\sum\limits_{k = 0}^{\left\lfloor {{k_{{\rm{th}}}}\left( m \right)} \right\rfloor } {\binom{k_{\max}}{k}{p_{\rm{tri}}}^k{{\left( {1 - {p_{\rm{tri}}}} \right)}^{{k_{\max }} - k}}} } \right.} \\
\left. { + \sum\limits_{k = \left\lceil {{k_{{\rm{th}}}}\left( m \right)} \right\rceil }^{{k_{\max }}} {\binom{k_{\max}}{k}{p_{\rm{tri}}}^k{{\left( {1 - {p_{\rm{tri}}}}\right)}^{{k_{\max }} - k}}} } \right\}
\end{dmath}
where $\left\lfloor \cdot  \right\rfloor$ and $\left\lceil  \cdot  \right\rceil $ are the floor and ceiling functions. denotes the decision threshold between symbols “$m-1$” and “$m$”, derived via the maximum likelihood (ML) criterion \cite{ref13}:
\begin{align}\label{eq17}
{k_{{\rm{th}}}}\left( m \right) = \frac{{{k_{\max }}\ln \left[ {\frac{{1 - {p_{\rm{tri}}}\left( {\alpha ,m} \right)}}{{1 - {p_{\rm{tri}}}\left( {\alpha ,m + 1} \right)}}} \right]}}{{\ln \left[ {\frac{{{p_{\rm{tri}}}\left( {\alpha ,m + 1} \right)\left( {1 - {p_{\rm{tri}}}\left( {\alpha ,m} \right)} \right)}}{{{p_{\rm{tri}}}\left( {\alpha ,m} \right)\left( {1 - {p_{\rm{tri}}}\left( {\alpha ,m + 1} \right)} \right)}}} \right]}}   
\end{align}

In practical systems, SER serves as an intuitive performance indicator, especially in the absence of forward error correction, and is used herein to quantify the improvement afforded by the AAC technique in Section. IV.

\section{Numerical Results and Discussions}
This section presents numerical simulations evaluating the proposed AAC technique in SPAD-based OWC systems. We compare two variants: the achievable rate optimization algorithm Alg. \ref{alg_1}, which exhaustively enumerates all PMFs, and the low-complexity alternative Alg. \ref{alg_2}, based on the average trigger probability criterion. The achievable rate is computed analytically using (\ref{eq6}), and the SER is obtained from (\ref{eq16}). The performance of the OWC system equipped with the AAC module is benchmarked against a traditional system without AAC.

In simulations, the dead time is set between one to tens of nanoseconds, consistent with commercially available SPAD devices. The signal constellation for 4‑PAM is designed using the square-root signaling method, which is well-suited for SDN channels \cite{ref34, ref35, ref36}. Under this scheme, the signal constellation points are given by $\left\{ {0,{\frac{1}{9}}{\lambda _{\rm{s}}},{\frac{4}{9}}{\lambda _{\rm{s}}},{\lambda _{\rm{s}}}} \right\}$.

\begin{figure*}[!t]
\centering
\subfloat[Optimal attenuation factor]{\includegraphics[width=3.3in]{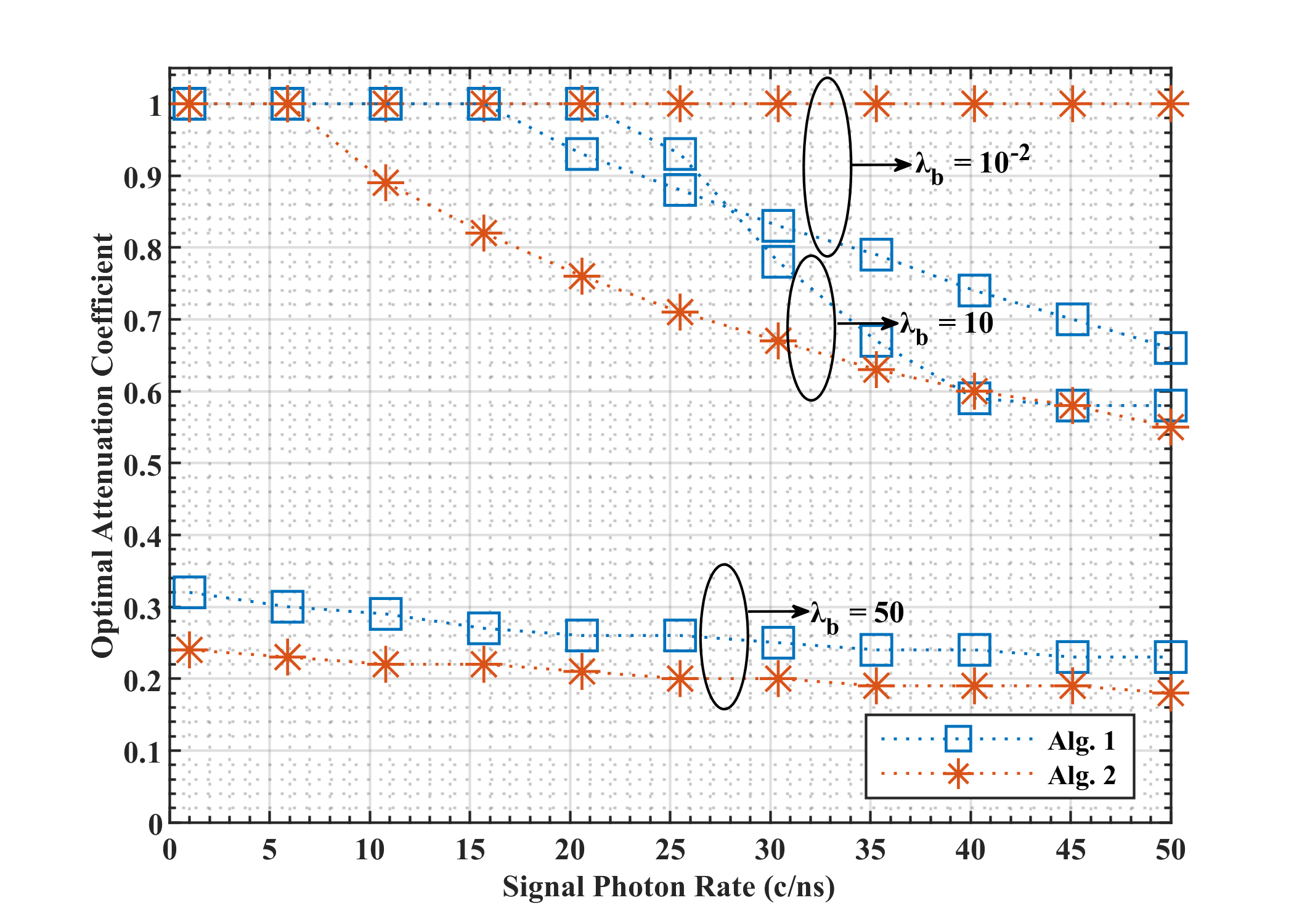}\label{fig_4_1}}
\hfil
\subfloat[Achievable rate]{\includegraphics[width=3.3in]{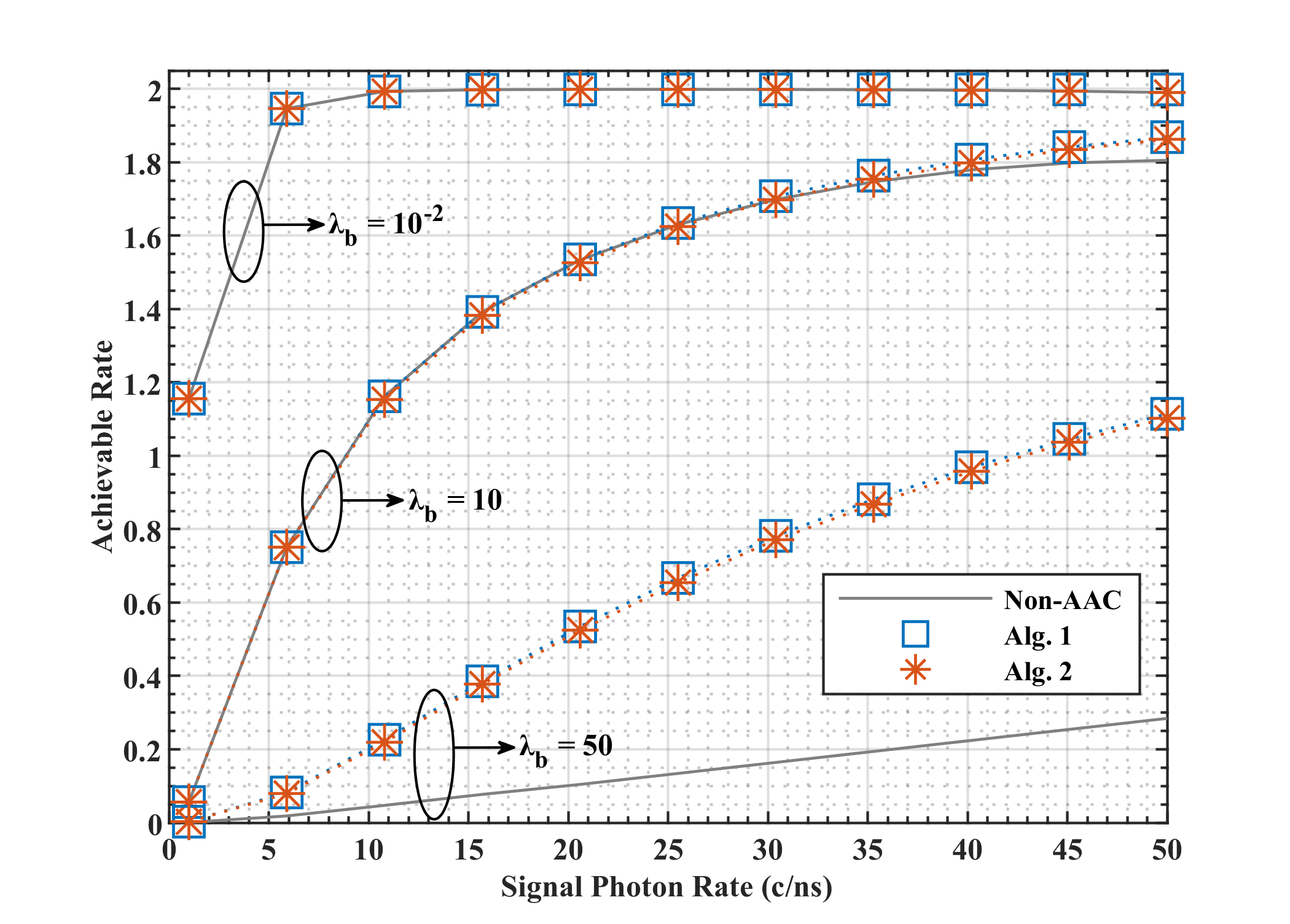}\label{fig_4_2}}
\hfil
\subfloat[Average trigger probability]{\includegraphics[width=3.3in]{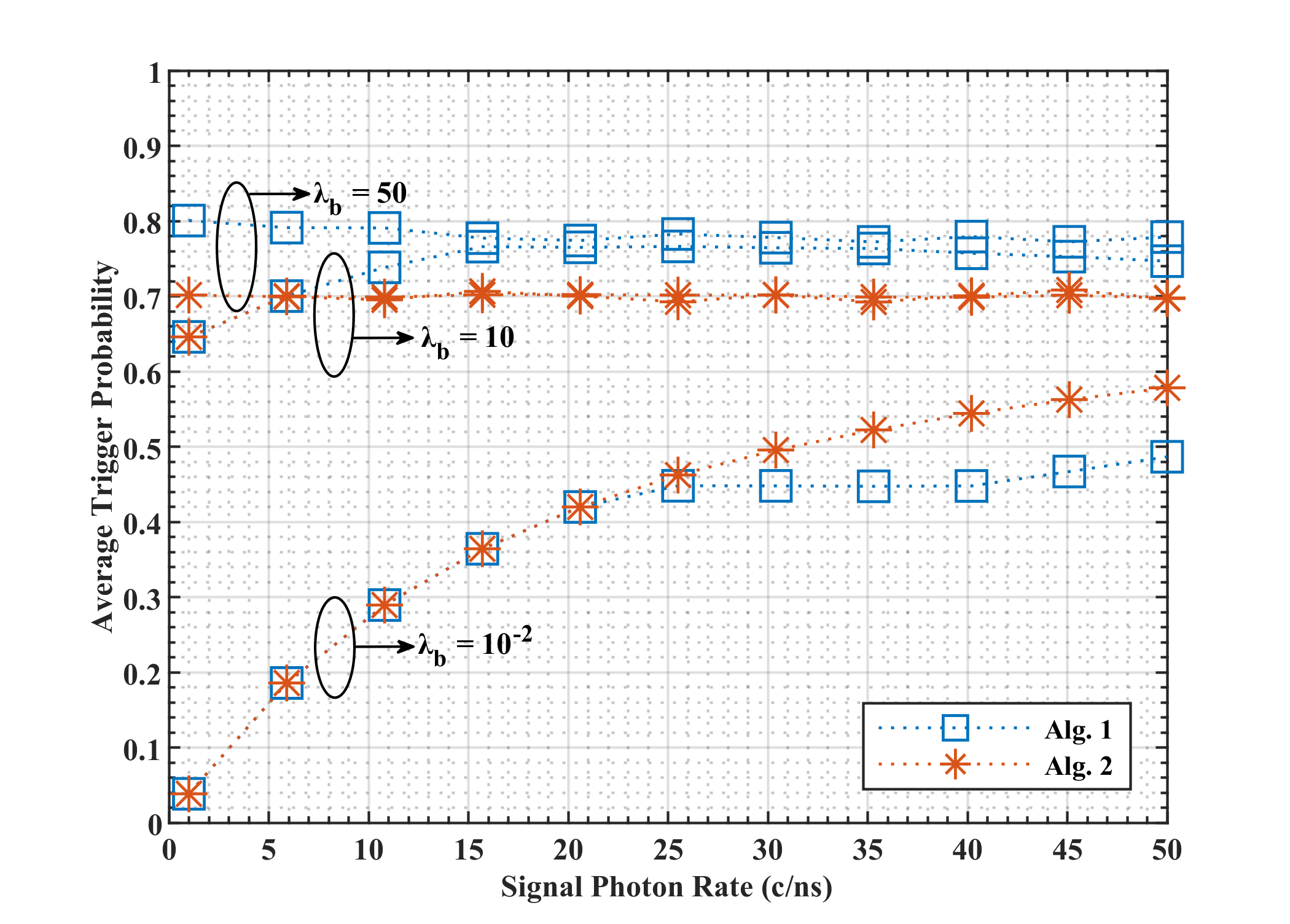}\label{fig_4_3}}
\hfil
\subfloat[SER]{\includegraphics[width=3.3in]{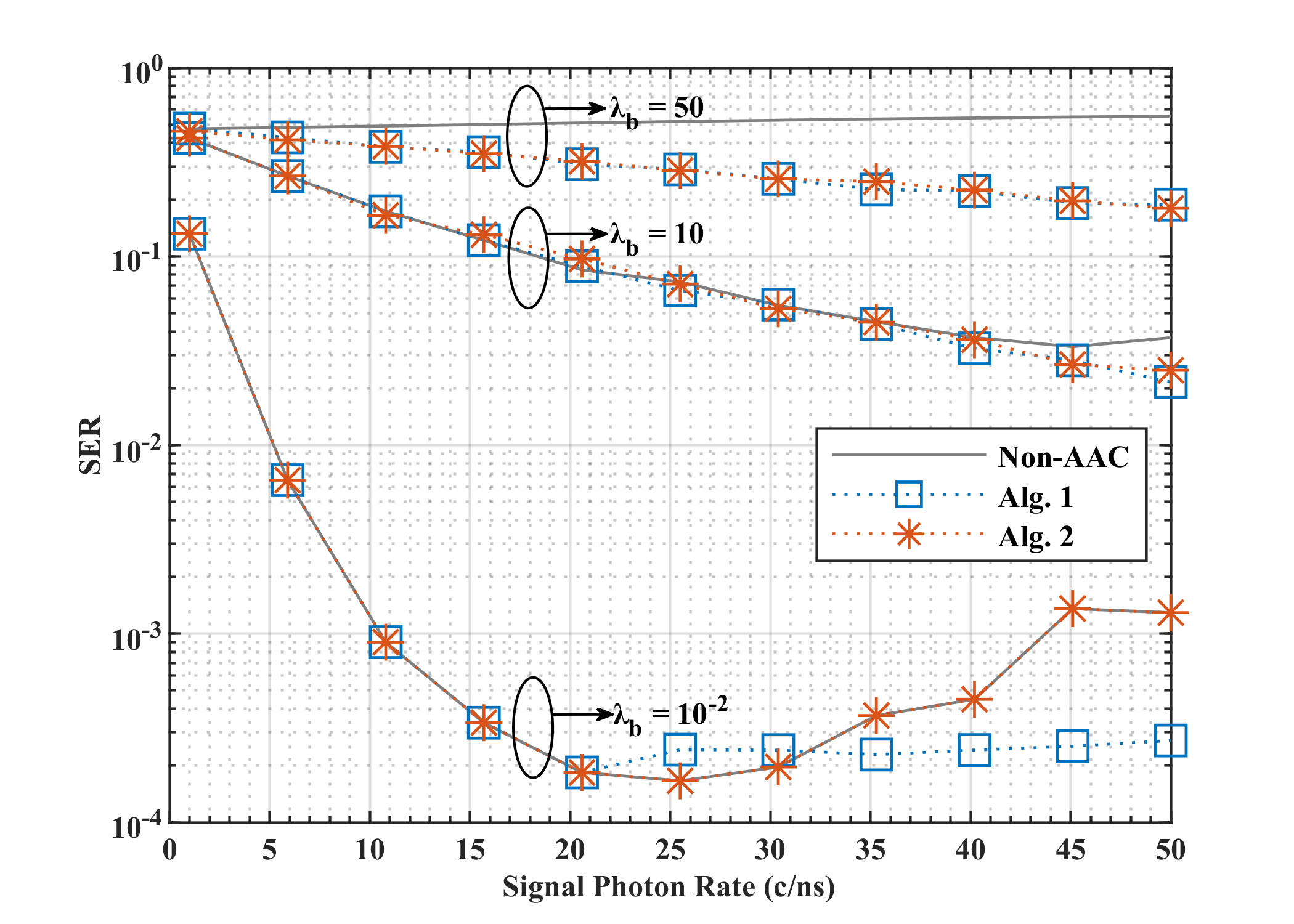}\label{fig_4_4}} 
\caption{Performance metric comparisons (with vs. without AAC) as a function of signal photon rate for different background radiations (${k_{\max}} = 100$).}
\label{fig_4}
\end{figure*}

Fig. \ref{fig_4_1} illustrates the optimal attenuation factor ${\alpha _{{\rm{opt}}}}$ as a function of the signal photon rate ${\lambda _{\rm{s}}}$ under different background radiation levels. At weak background radiation (${\lambda _{\rm{b}}} = 0.1$), the SPAD operates in the linear region without saturation; hence, ${\alpha _{{\rm{opt}}}} = 1$ and the AAC module remains inactive. Under stronger background conditions (${\lambda _{\rm{b}}} = 10,50$), however, the SPAD experiences saturation. Here, ${\alpha _{{\rm{opt}}}}$ decreases rapidly with increasing  to maintain the incident photon flux within the SPAD’s linear dynamic range. Moreover, a higher background level necessitates a lower ${\alpha _{{\rm{opt}}}}$ to effectively attenuate the total incident optical power.

Fig. \ref{fig_4_2} demonstrates that the achievable rate with the AAC technique significantly outperforms that of the non-optimized system. Under severe background radiation (${\lambda _{\rm{b}}} = 50$), the achievable rate of the non-optimized system shows negligible improvement with increasing ${\lambda _{\rm{s}}}$, whereas systems using AAC exhibit rapid achievable rate growth. The achievable rate from Alg. \ref{alg_2} is slightly lower than that from Alg. \ref{alg_1}, as the former provides a near-optimal solution while the latter yields the theoretically optimal value.

Fig. \ref{fig_4_3} compares the average trigger probability obtained via Alg. \ref{alg_1} and Alg. \ref{alg_2}. Under weak background radiation, both algorithms yield nearly identical average trigger probability. However, under strong background conditions, Alg. \ref{alg_1} produces a higher average trigger probability than Alg. \ref{alg_2}, since the latter enforces a fixed benchmark of 0.7 across all scenarios.

Fig. \ref{fig_4_4} shows the corresponding SER performance. Under strong background conditions, the non-optimized SER remains almost unchanged with increasing ${\lambda _{\rm{s}}}$. In contrast, both AAC algorithms drive the SER down rapidly. At ${\lambda _{\rm{s}}} = 50$, the optimized SER is nearly one order of magnitude lower than that of the non-optimized system, confirming the effectiveness of AAC in improving error performance.

\begin{figure*}[!t]
\centering
\subfloat[Optimal attenuation factor]{\includegraphics[width=3.3in]{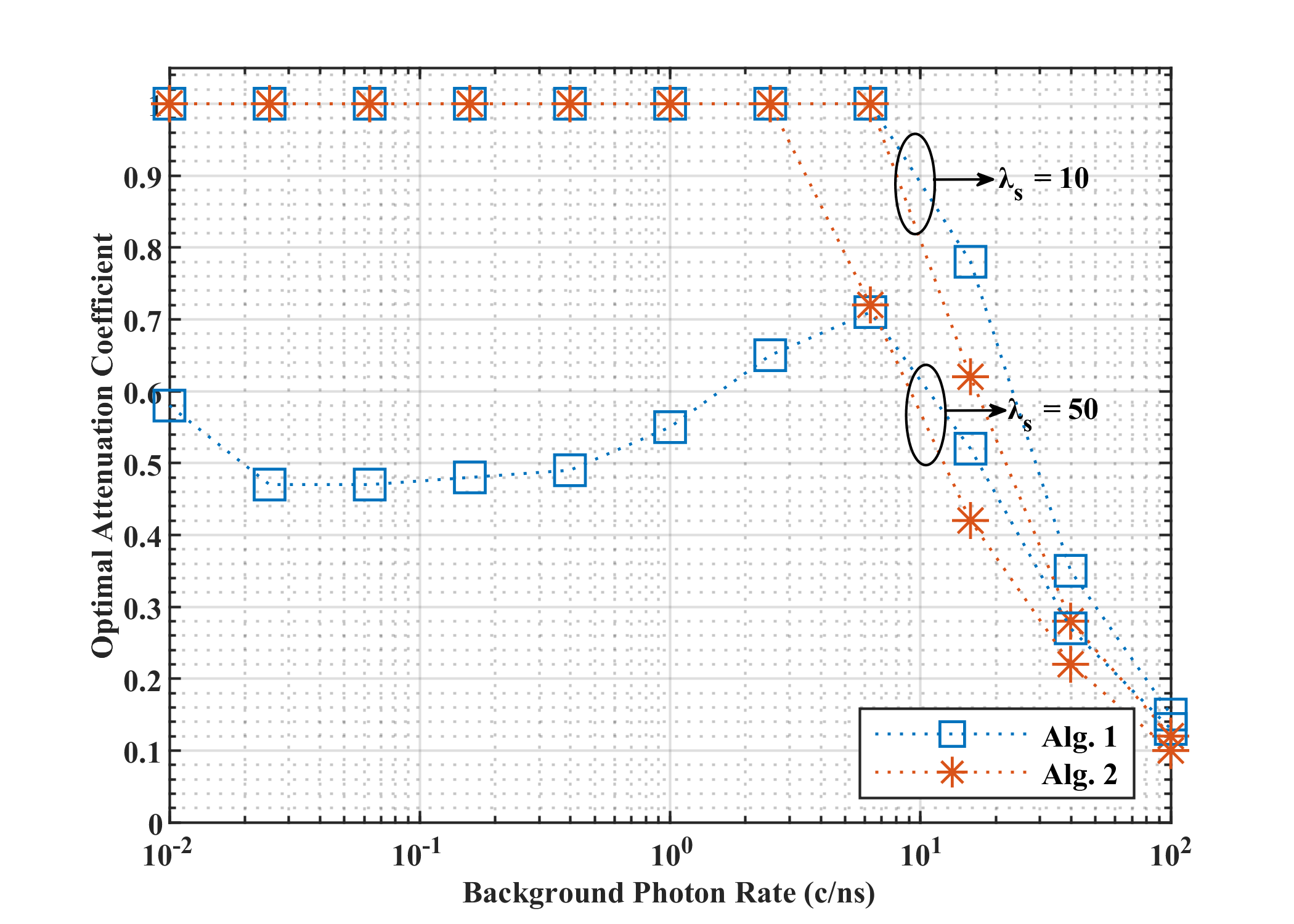}\label{fig_5_1}}
\hfil
\subfloat[Achievable rate]{\includegraphics[width=3.3in]{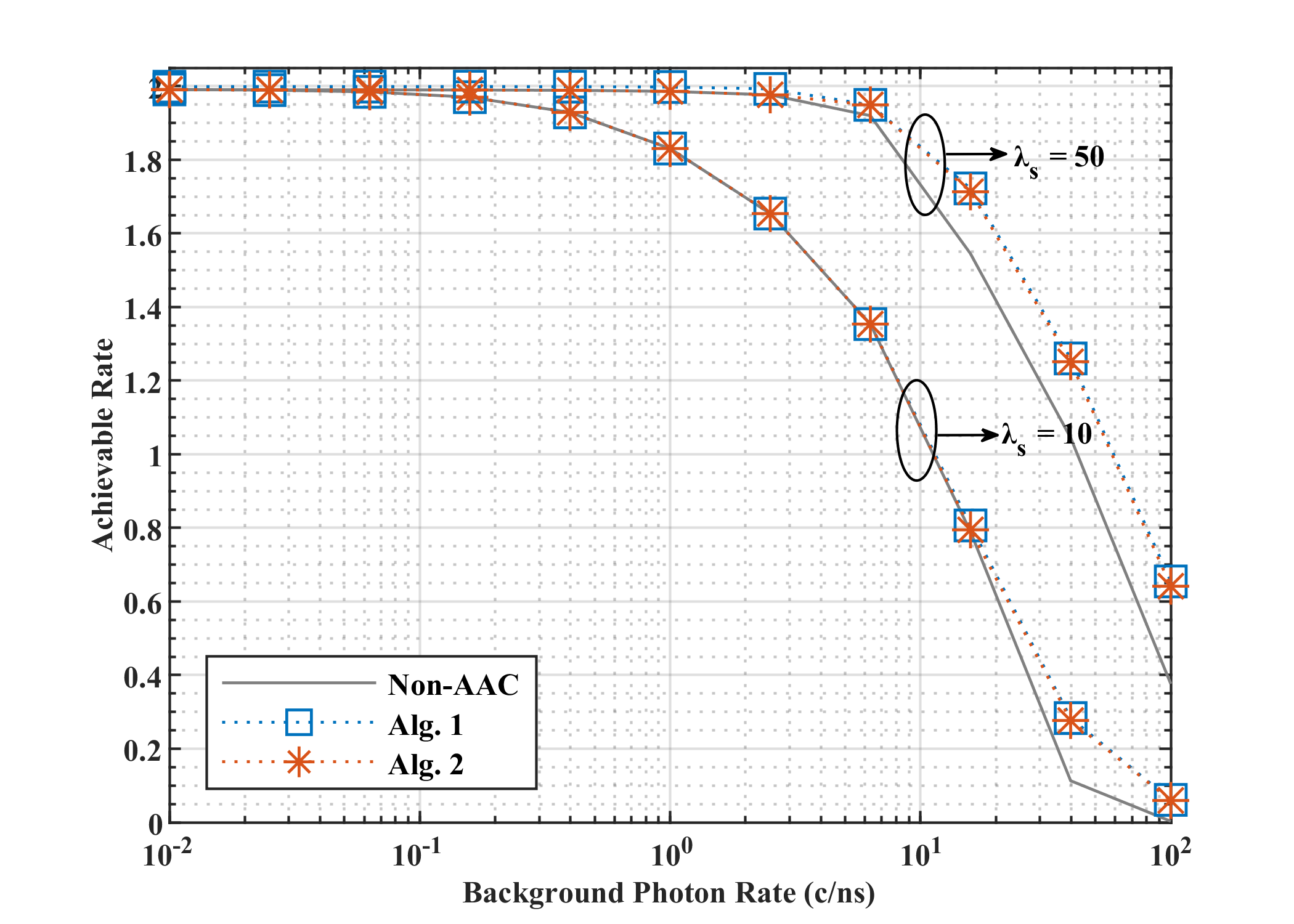}\label{fig_5_2}}
\hfil
\subfloat[Average trigger probability]{\includegraphics[width=3.3in]{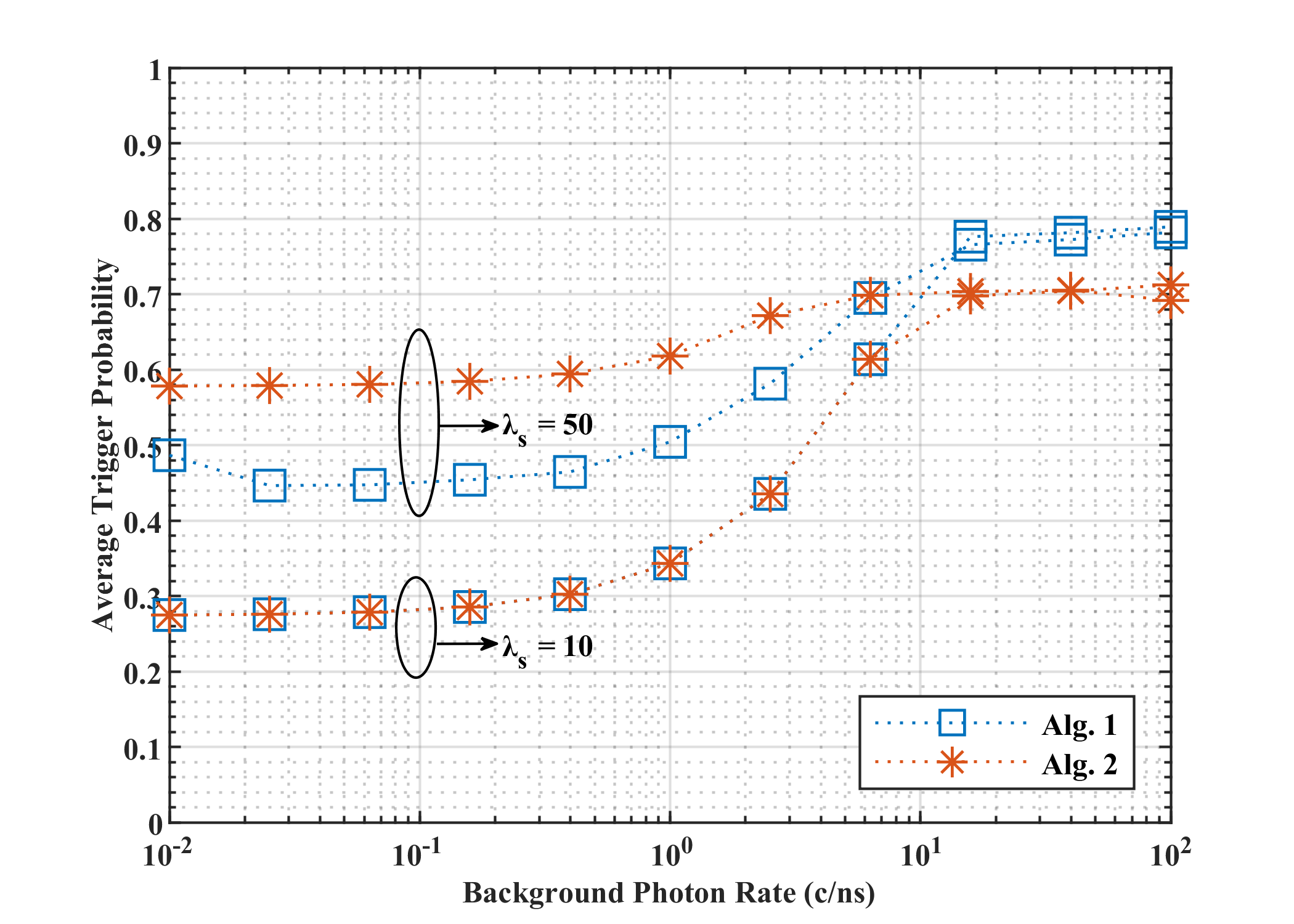}\label{fig_5_3}}
\hfil
\subfloat[SER]{\includegraphics[width=3.3in]{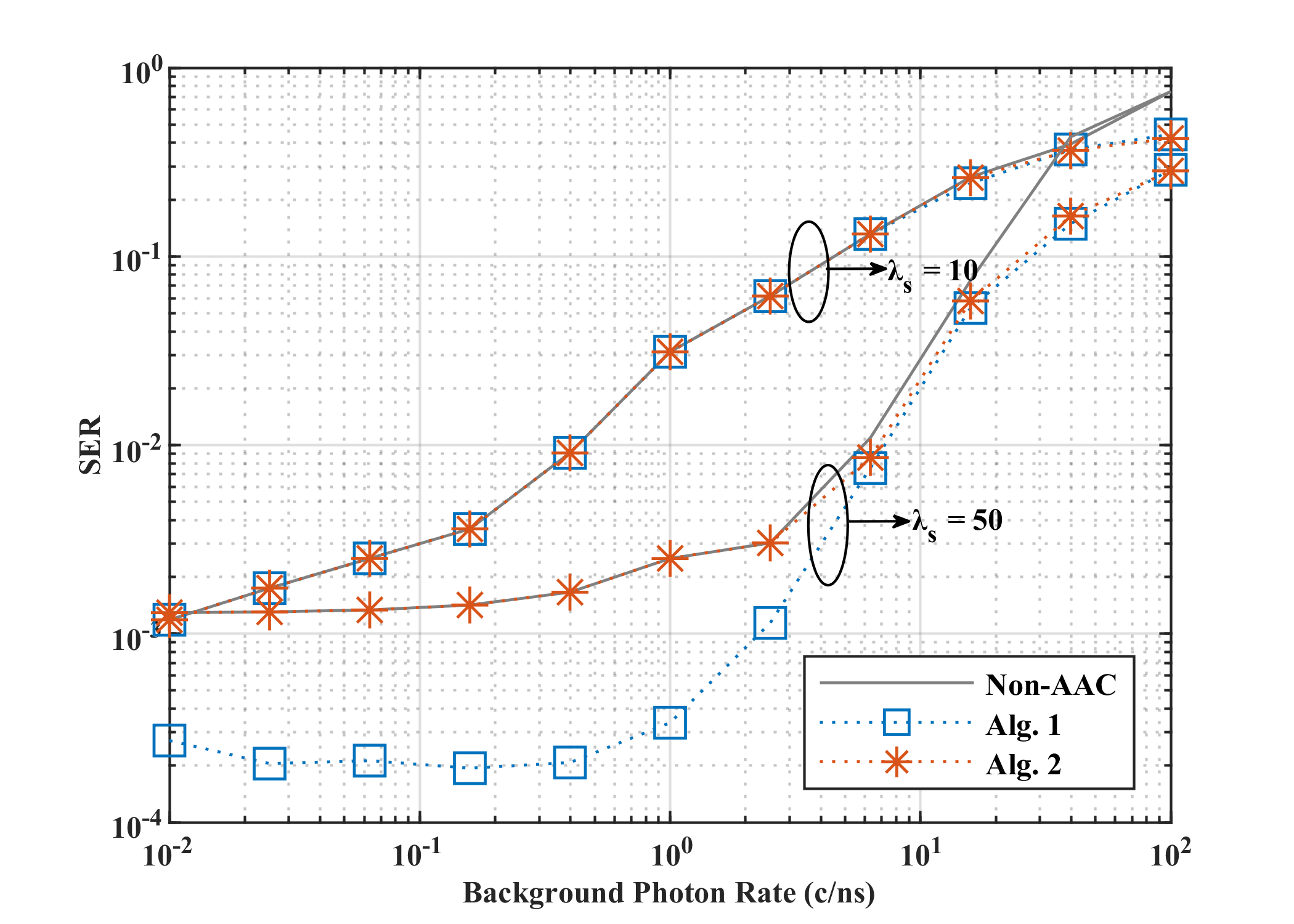}\label{fig_5_4}}
\caption{Performance metric comparisons (with vs. without AAC) as a function of background photon rate for different incident signal intensities (${k_{\max}} = 100$).}
\label{fig_5}
\end{figure*}

As shown in Fig. \ref{fig_5_1}, the optimal attenuation factor ${\alpha _{{\rm{opt}}}}$ remains constant in the weak background range (non-saturated regime) and then decreases rapidly in the strong background range (saturated regime), consistent with the results in Fig. \ref{fig_4_1}. This behavior occurs because excessive background photons saturate the photon counts across all symbols, degrading both achievable rate and SER.

Fig. \ref{fig_5_3} indicates that the average trigger probability from both algorithms is similar in most cases. However, under high signal intensity with weak background radiation, Alg. \ref{alg_2} results in a higher average trigger probability than Alg. \ref{alg_1}, slightly degrading system performance due to the fixed trigger probability benchmark.

Fig. \ref{fig_5_2} and \ref{fig_5_4} show that as ${\lambda _{\rm{b}}}$ increases, the improvements in achievable rate and SER become more pronounced. The AAC technique effectively mitigates saturation under strong background radiation. In most scenarios, Alg. \ref{alg_2} performs comparably to Alg. \ref{alg_1}. However, under high signal intensity (${\lambda _{\rm{s}}} = 50$) and negligible background (${\lambda _{\rm{b}}} \le 1$), Alg. \ref{alg_1} improves the SER by approximately one order of magnitude over Alg. \ref{alg_2}. This minor performance degradation remains acceptable, especially given that Alg. \ref{alg_2} closely matches Alg. \ref{alg_1} under strong background conditions where AAC is most critical. Therefore, Alg. \ref{alg_2} achieves performance comparable to that of Alg. \ref{alg_1} while significantly reducing the computational burden. 

\begin{figure*}[!t]
\centering
\subfloat[Optimal attenuation factor]{\includegraphics[width=3.3in]{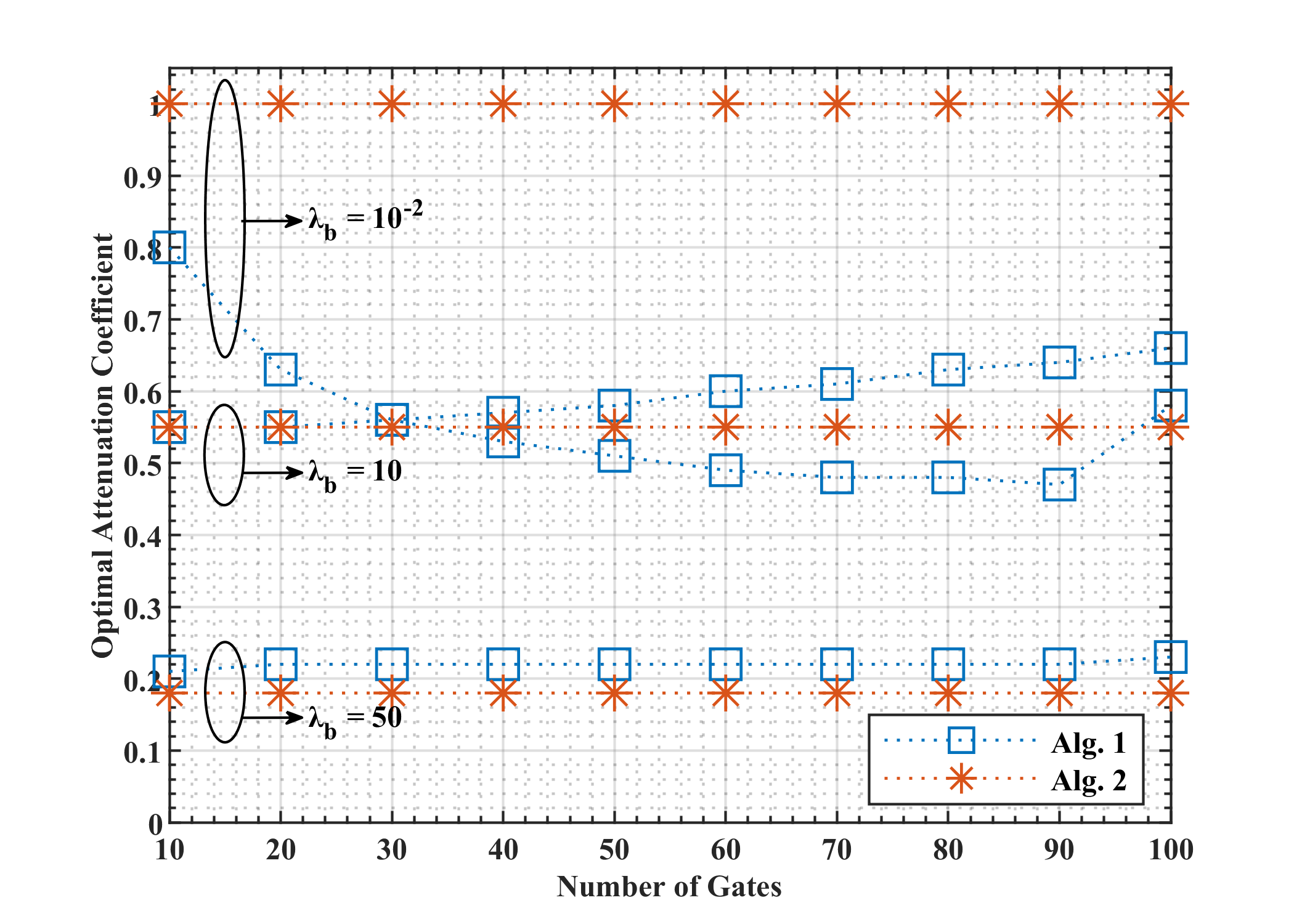}\label{fig_6_1}}
\hfil
\subfloat[Achievable rate]{\includegraphics[width=3.3in]{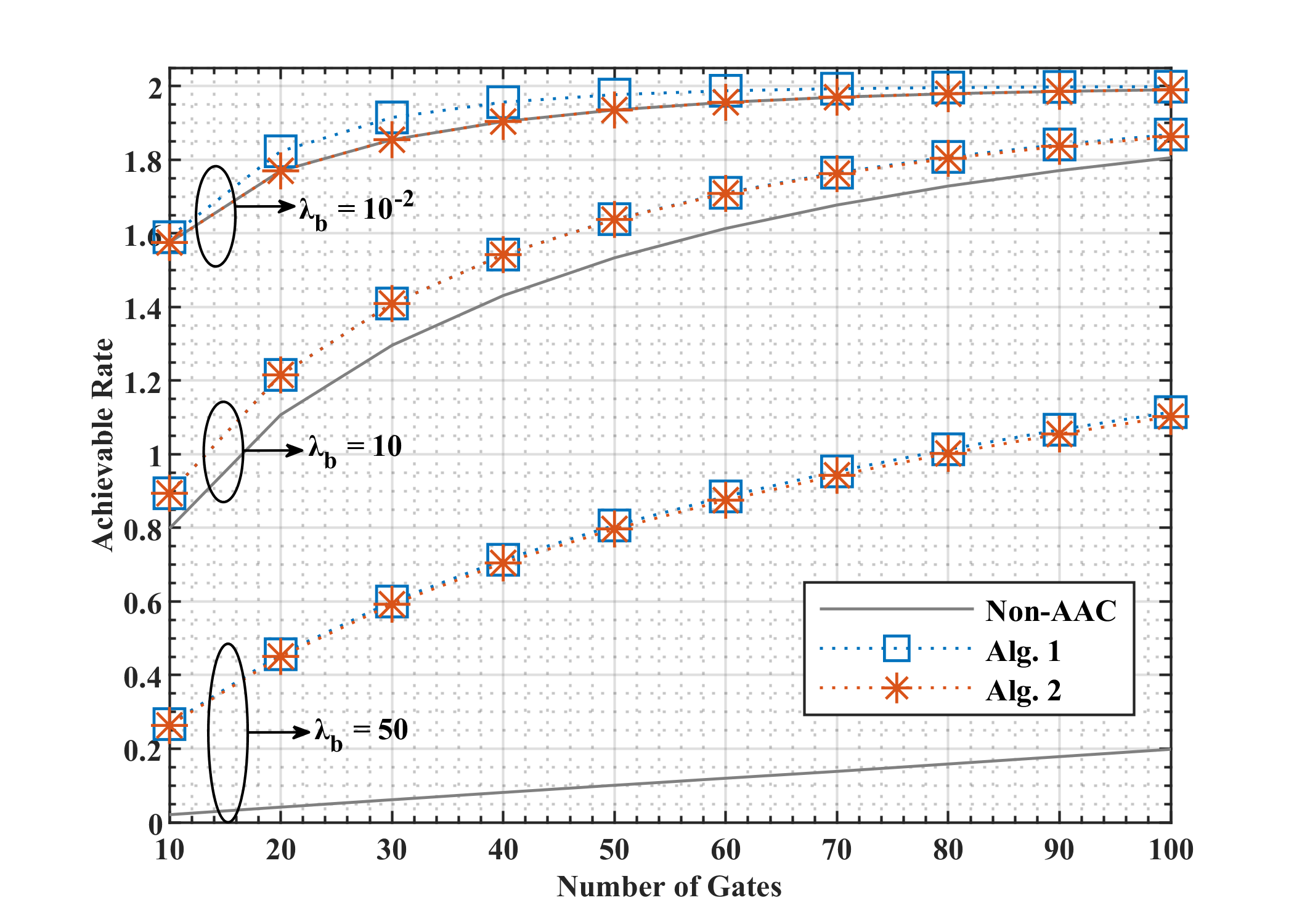}\label{fig_6_2}}
\hfil
\subfloat[Average trigger probability]{\includegraphics[width=3.3in]{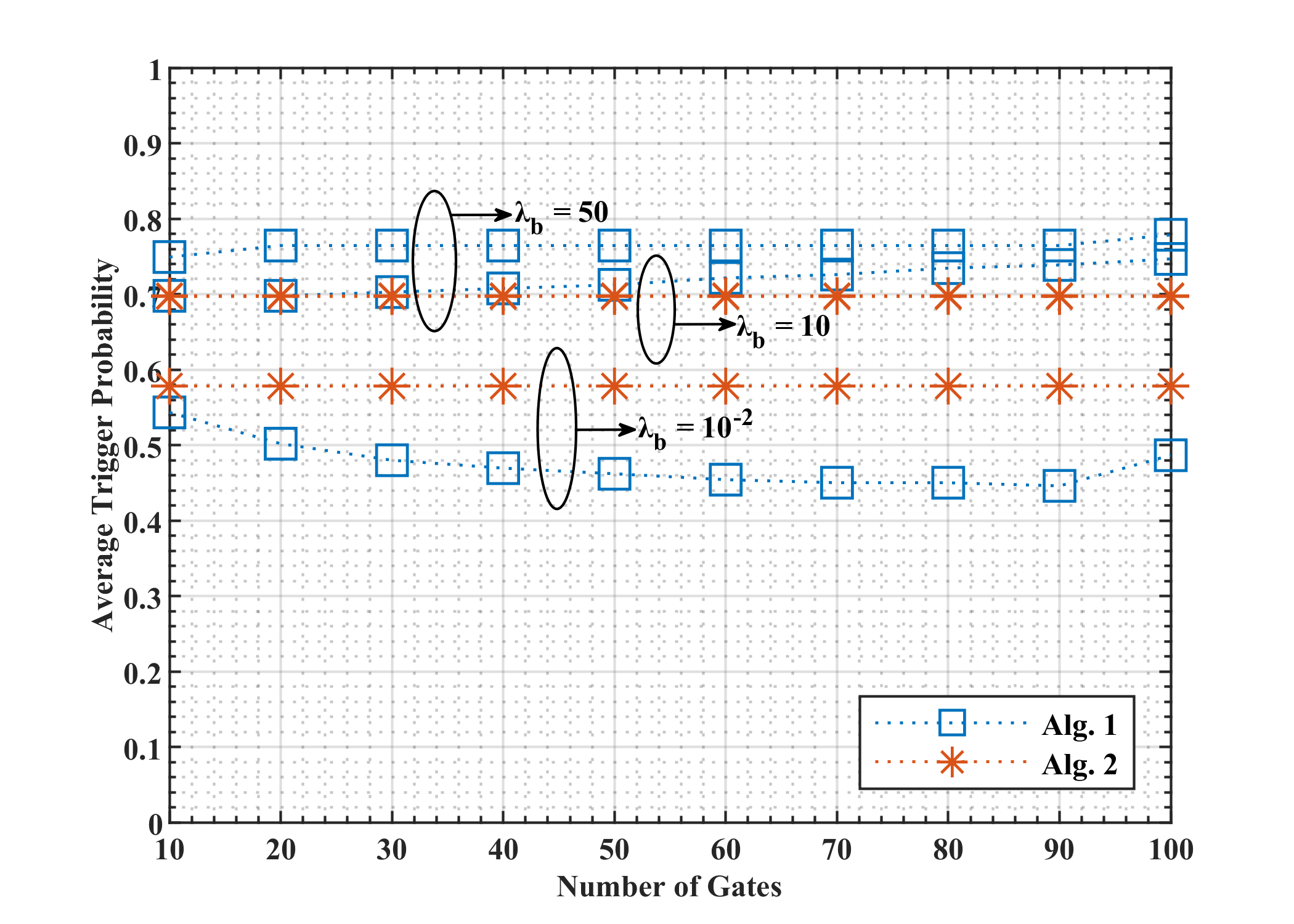}\label{fig_6_3}}
\hfil
\subfloat[SER]{\includegraphics[width=3.3in]{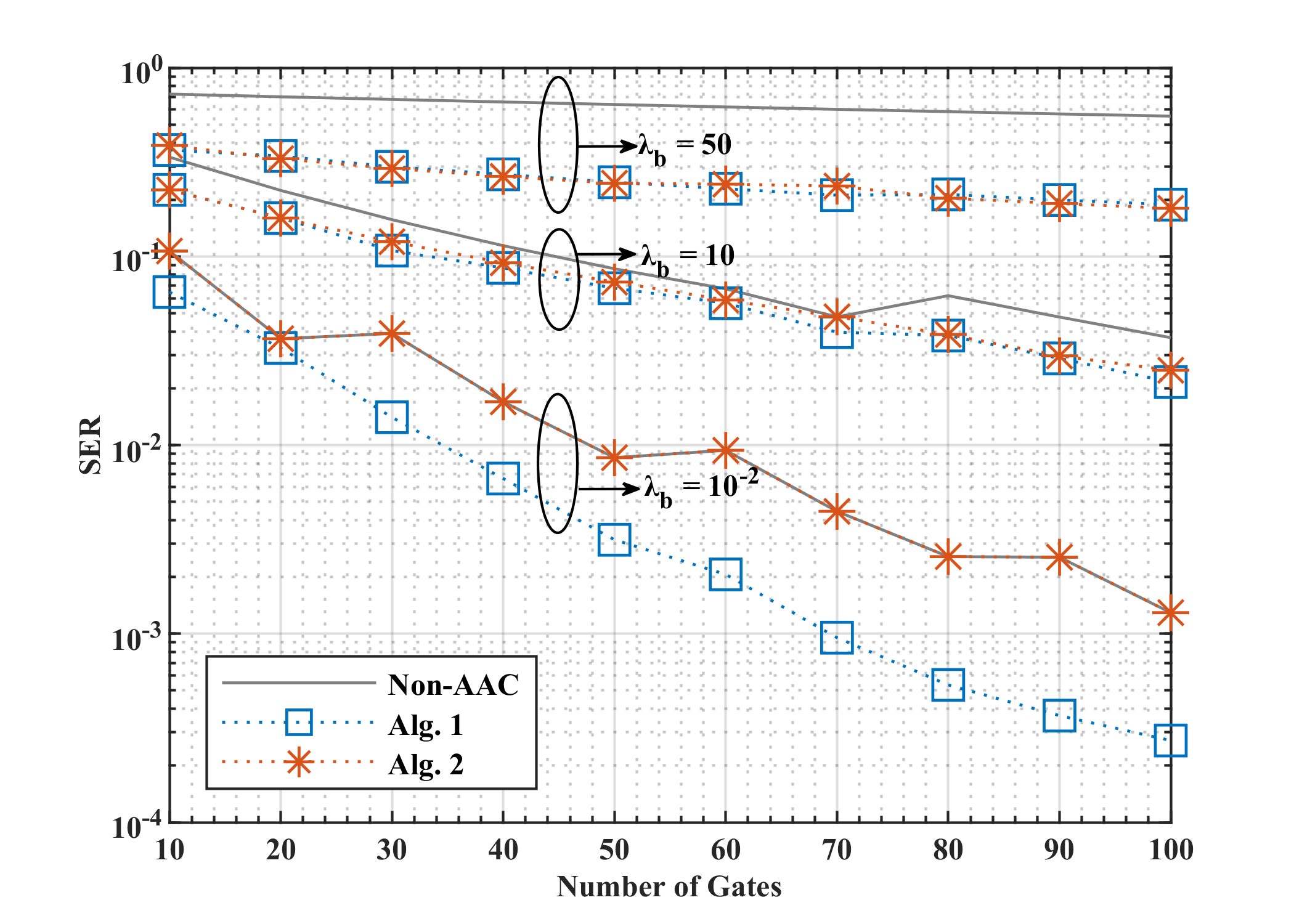}\label{fig_6_4}}
\caption{Performance metric comparisons (with vs. without AAC) as a function of number of gates for different background radiations (${\lambda _{\rm{s}}} = 50$).}
\label{fig_6}
\end{figure*}

Fig. \ref{fig_6_1} shows that the optimal attenuation factor ${\alpha _{{\rm{opt}}}}$ remains relatively stable as the number of gates increases. Combined with earlier results, this confirms that ${\alpha _{{\rm{opt}}}}$ is primarily determined by the signal and background photon rates, rather than the number of gates.

As seen in Fig. \ref{fig_6_2} and \ref{fig_6_4}, the achievable rate and SER improvements are strongly influenced by the number of gates. In SPAD-based systems, communication performance depends heavily on the number of gates, which shapes the PMF profiles. The AAC technique helps maintain SER near the lower bound even under high incident photon flux. As the number of gates increases, the SER decreases exponentially. With AAC maintaining optimal incident intensity, the SER improvement can reach up to one order of magnitude.

In summary, the proposed AAC technique significantly enhances the performance of SPAD-based OWC systems under both weak and strong background radiation. Alg. \ref{alg_2} achieves performance comparable to that of Alg. \ref{alg_1} while reducing computational complexity by over two orders of magnitude, making it highly suitable for implementation in low-power terminals with modest performance trade-offs.

\section{Conclusion}
In conclusion, we have developed a comprehensive analytical framework to model the photon-counting saturation behavior in practical SPAD-based receivers, incorporating the effects of dead time and lack of PNR. An AAC technique has been proposed to mitigate receiver saturation and maintain the incident optical signal at an optimal level. The proposed AAC technique effectively extends the dynamic range of photon-counting receivers in the presence of strong background radiation or high signal power. Numerical results confirm that the AAC module substantially improves both the achievable rate and SER across a wide range of operating conditions. With its low-complexity implementation, the AAC module holds great potential as an essential optical intensity control component in future OWC systems with photon-counting receivers.

\vfill


\begin{thebibliography}{1}
\bibliographystyle{IEEEtran}

\bibitem{ref1}
G. Park, V. Mai, H. Lee, S. Lee, and H. Kim, "Free-Space Optical Communication Technologies for Next-Generation Cellular Wireless Communications," \textit{IEEE Communications Magazine}, vol. 62, no. 3, pp. 24-30, 2024.

\bibitem{ref2}
J. Zhang et al., "Long-term and Real-time High-speed Underwater Wireless Optical Communications in Deep Sea," \textit{IEEE Communications Magazine}, vol. 62, no. 3, pp. 96-101, 2023.

\bibitem{ref3}
Zimmermann, "APD and SPAD Receivers : Invited Paper," \textit{in 2019 15th International Conference on Telecommunications (ConTEL)}, 2019, pp. 1-5.

\bibitem{ref4}
X. Jiang, M. A. Itzler, R. Ben-Michael, and K. Slomkowski, "InGaAsP–InP Avalanche Photodiodes for Single Photon Detection," \textit{IEEE Journal of Selected Topics in Quantum Electronics}, vol. 13, no. 4, pp. 895-905, 2007.

\bibitem{ref5}
R. J. Cesarone, D. S. Abraham, S. Shambayati, and J. Rush, "Deep-space optical communications," \textit{in 2011 International Conference on Space Optical Systems and Applications (ICSOS)}, 2011, pp. 410-423.

\bibitem{ref6}
M. F. Ali, D. N. K. Jayakody, Y. A. Chursin, S. Affes, and S. Dmitry, "Recent Advances and Future Directions on Underwater Wireless Communications," \textit{Archives of Computational Methods in Engineering}, vol. 27, no. 5, pp. 1379-1412, 2020.

\bibitem{ref7}
D. Chitnis and S. Collins, "A SPAD-Based Photon Detecting System for Optical Communications," \textit{Journal of Lightwave Technology}, vol. 32, no. 10, pp. 2028-2034, 2014.

\bibitem{ref8}
Y. Zhang, X. Wang, Z. Du, Y. Gao, and J. Xu, "A High-Speed Photon-Counting UWOC System With Multiple Channels to Suppress the Randomness in Detection," \textit{Journal of Lightwave Technology}, vol. 42, no. 20, pp. 7185-7192, 2024.

\bibitem{ref9}
R. B. Benjamin et al., "photon counting camera for the NASA deep space optical communication demonstration on the PSYCHE mission," \textit{in Proc.SPIE}, 2019, vol. 10978, p. 1097809.

\bibitem{ref10}
F. Cucchietti, A. Gallesio, M. Liberatore and A. Piccirillo, "Impact of Photodetector Defects on Optical System Performance," \textit{IEEE Journal on Selected Areas in Communications}, vol. 4, no. 7, pp. 1121-1130, 1986.

\bibitem{ref11}
J. P. Donnelly et al., "Design Considerations for 1.06 $\mu$m InGaAsP–InP Geiger-Mode Avalanche Photodiodes," \textit{IEEE Journal of Quantum Electronics}, vol. 42, no. 8, pp. 797-809, 2006.

\bibitem{ref12}
M. A. Itzler et al., "Single photon avalanche diodes (SPADs) for 1.5 $\mu$m photon counting applications," \textit{Journal of Modern Optics}, vol. 54, no. 2-3, pp. 283-304, 2007.

\bibitem{ref13}
C. Wang et al., “Performance analysis of photon-limited free-space optical communications with practical photon-counting receivers,” \textit{Optics Express}, vol. 33, no. 5, pp. 10741-10758, 2025.

\bibitem{ref14}
M. Teich and B. Cantor, "Information, error, and imaging in deadtime-perturbed doubly stochastic Poisson counting systems," \textit{IEEE Journal of Quantum Electronics}, vol. 14, no. 12, pp. 993-1003, 1978.

\bibitem{ref15}
C.-C. Chen, "Effect of Detector Dead Time on the Performance of Optical Direct-Detection Communication Links," \textit{JPL TDA Progress Report}, vol. 46-53, pp. 146-154, 1988.

\bibitem{ref16}
X. Li et al., "Enhanced photon communication through Bayesian estimation with an SNSPD array," \textit{Photonics Research}, vol. 8, no. 5, pp. 637-641, 2020.

\bibitem{ref17}
X. Li, D. Li, C. Tang, and S. Liu, "Fully-Integrated SPAD-Based Receiver With Nanosecond Dead Time for Optical Wireless Communication," \textit{Journal of Lightwave Technology}, vol. 41, no. 2, pp. 653-661, 2023.

\bibitem{ref18}
J. Li, D. Ye, K. Fu, L. Wang, J. Piao, and Y. Wang, "Single-photon detection for MIMO underwater wireless optical communication enabled by arrayed LEDs and SiPMs," \textit{Optics Express}, vol. 29, no. 16, pp. 25922-25944, 2021.

\bibitem{ref19}
S. Hu, L. Mi, T. Zhou, and W. Chen, "35.88 attenuation lengths and 3.32 bits/photon underwater optical wireless communication based on photon-counting receiver with 256-PPM," \textit{Optics Express}, vol. 26, no. 17, pp. 21685-21699, 2018.

\bibitem{ref20}
Q. Yan, Z. Li, Z. Hong, T. Zhan, and Y. Wang, "Photon-Counting Underwater Wireless Optical Communication by Recovering Clock and Data From Discrete Single Photon Pulses," \textit{ IEEE Photonics Journal}, vol. 11, no. 5, pp. 1-15, 2019.

\bibitem{ref21}
Q. Yan, M. Wang, W. Dai, and Y. Wang, "Synchronization scheme of photon-counting underwater optical wireless communication based on PPM," \textit{Optics Communications}, vol. 495, p. 127024, 2021.

\bibitem{ref22}
C. Wang et al., "Performance analysis of photon-limited PPM-FSO communications: SER under afterpulsing effect in practical photon-counting receivers," \textit{Optics Express}, vol. 33, no. 18, pp. 38352-38363, 2025.

\bibitem{ref23}
G. Wen, J. Huang, J. Dai, L. Zhang, and J. Wang, "Performance analysis optimization and experimental verification of a photon-counting communication system based on non-photon-number-resolution detectors," \textit{Optics Communications}, vol. 468, p. 125771, 2020.

\bibitem{ref24}
G. Wen, J. Huang, L. Zhang, C. Li, T. Wen, and J. Wang, "A High-Speed and High-Sensitivity Photon-Counting Communication System Based on Multichannel SPAD Detection," \textit{ IEEE Photonics Journal}, vol. 13, no. 2, pp. 1-10, 2021.

\bibitem{ref25}
J. Huang, C. Li, J. Dai, R. Shu, L. Zhang, and J. Wang, "Real-Time and High-Speed Underwater Photon-Counting Communication Based on SPAD and PPM Symbol Synchronization," \textit{ IEEE Photonics Journal}, vol. 13, no. 5, pp. 1-9, 2021.

\bibitem{ref26}
S. Huang and M. Safari, "Time-Gated photon counting Receivers for Optical Wireless Communication," \textit{Journal of Lightwave Technology}, vol. 39, no. 22, pp. 7113-7123, 2021.

\bibitem{ref27}
Y. Mu et al., "Time-coordinated SPAD-based receiver for high-speed optical wireless communication," \textit{Optics Communications}, vol. 526, p. 128706, 2023.

\bibitem{ref28}
L. Zhang et al., "A Comparison of APD- and SPAD-Based Receivers for Visible Light Communications," \textit{Journal of Lightwave Technology}, vol. 36, no. 12, pp. 2435-2442, 2018.

\bibitem{ref29}
S. Huang and M. Safari, "Hybrid SPAD/PD Receiver for Reliable Free-Space Optical Communication," \textit{IEEE Open Journal of the Communications Society}, vol. 1, pp. 1364-1373, 2020.

\bibitem{ref30}
S. Huang et al., "Single-photon counting Receivers for Optical Wireless Communications in Future 6G Networks," \textit{IEEE Communications Magazine}, vol. 62, no. 3, pp. 54-60, 2024.

\bibitem{ref31}
C. Wang et al., " BER improvement in SPAD-based photon-counting optical communication system by using automatic attenuation control technique," \textit{Optics Letters}, vol. 47, no. 8, pp. 1956-1959, 2022.

\bibitem{ref32}
B. I. Cantor and M. C. Teich, "Dead-time-corrected photocounting distributions for laser radiation," \textit{Journal of the Optical Society of America}, vol. 65, no. 7, pp. 786-791, 1975.

\bibitem{ref33}
R. M. Gagliardi and S. Karp, \textit{Optical Communications}. 2nd ed. New York, NY, USA: Wiley, 1995.

\bibitem{ref34}
Q. Gao, S. Hu, C. Gong, and Z. Xu, "Modulation Designs for Visible Light Communications With Signal-Dependent Noise," \textit{Journal of Lightwave Technology}, vol. 34, no. 23, pp. 5516-5525, 2016.

\bibitem{ref35}
L. H. Si-Ma, H. Y. Yu, J. Zhang, G. Xin, C. Wang, and R. H. Chen, "Multidimensional Multilayer Modulation for Broadcast UVLC With Photon Detectors," \textit{IEEE Transactions on Vehicular Technology}, vol. 69, no. 6, pp. 6437-6451, 2020.

\bibitem{ref36}
Q. Gao, K. Qaraqe, and E. Serpedin, "Improving the Modulation Designs for Visible Light Communications with Signal-Dependent Noise," \textit{IEEE Communications Magazine}, vol. 58, no. 5, pp. 26-32, 2020.

\end{thebibliography}
\end{document}